\documentclass[
 reprint,
 amsmath,amssymb,
 aps,
]{revtex4-2}

\usepackage[dvipdfmx]{graphicx}
\usepackage{dcolumn}
\usepackage{bm}
\usepackage{color}
\usepackage[colorlinks=true, linkcolor=blue, citecolor=magenta, urlcolor=blue]{hyperref}
\usepackage{physics}
\usepackage{orcidlink}
\usepackage{tikz}
\usetikzlibrary{positioning, shapes.geometric}
\usetikzlibrary{arrows.meta}
\usetikzlibrary{calc}

\newcommand{\mapcube}[2]{\mathcal{M}_{{\rm ap}{#1}}^{3{#2}}}

\newcommand{\code}[1]{\textsc{#1}}
\newcommand{\codett}[1]{\texttt{#1}}

\begin{document}

\title{Cosmology from a joint analysis of second and third order shear statistics\\with Subaru Hyper Suprime-Cam Year 3 data}

\author{Sunao Sugiyama\orcidlink{0000-0003-1153-6735}}\email{ssunao@sas.upenn.edu}
\affiliation{Department of Physics and Astronomy, University of Pennsylvania, Philadelphia, PA 19104, USA}

\author{Rafael C. H. Gomes\orcidlink{0000-0002-3800-5662}}
\affiliation{Department of Physics and Astronomy, University of Pennsylvania, Philadelphia, PA 19104, USA}

\author{Bhuvnesh Jain \orcidlink{0000-0002-8220-3973}}
\affiliation{Department of Physics and Astronomy, University of Pennsylvania, Philadelphia, PA 19104, USA}

\date{\today}

\begin{abstract}
We present a joint cosmological analysis of the two-point correlation function and the aperture-mass skewness measured from the Year 3 data of the Hyper Suprime-Cam Subaru Strategic Program (HSC-Y3). The aperture-mass skewness is a compressed representation of three-point shear information, designed to capture non-Gaussian features while keeping the data vector computationally tractable. 
We find that including the aperture-mass skewness improves the $S_8$-$\Omega_m$ figure of merit by 80\% compared to the 2PCF-only case, primarily due to the breaking of degeneracies. Our joint analysis yields a constraint of $S_8=0.736\pm0.020$, which is slightly lower than the two-point-only result and increases the tension with Planck 2018 to 3.2$\sigma$ in the $S_8$-$\Omega_m$ plane. The two- and three-point statistics are found to be internally consistent across redshift bins and angular scales, and we detect no significant intrinsic alignment signal. 
We also explore extensions to the $w$CDM model and find no evidence for deviations from a cosmological constant.
This work demonstrates the feasibility and scientific value of incorporating  third-order shear statistics into weak lensing cosmology and provides a practical pathway for similar analyses in future Stage-IV surveys such as LSST, Euclid, and Roman.
\end{abstract}

\maketitle

\section{Introduction}
The current standard model of cosmology, the flat $\Lambda$CDM model, has been remarkably successful in describing a wide range of cosmological observations. However, in recent years, discrepancies have emerged between different probes. One of the most discussed tensions is the so-called $S_8$ tension, where weak lensing surveys tend to favor lower values of the parameter combination $S_8 \equiv \sigma_8 \sqrt{\Omega_{\rm m}/0.3}$ compared to those inferred from cosmic microwave background (CMB) observations, such as from the Planck satellite \citep{Collaboration.Zonca.2020}. This tension could be a statistical fluctuation, an indication of unknown systematics, or potentially a hint toward new physics beyond $\Lambda$CDM.

To address these questions, several ongoing and completed Stage-III weak lensing surveys -- including the Hyper Suprime-Cam Subaru Strategic Program \citep[HSC;][]{Aihara.Yuma.2018}, the Dark Energy Survey \citep[DES;][]{Collaboration.Zuntz.2016}, and the Kilo-Degree Survey \citep[KiDS;][]{Jong.consortiums.2012} -- have carried out deep and wide imaging programs. These surveys are now reaching their final data releases and delivering cosmological results with ever-increasing precision \cite{Heymans.Velander.2013, Hildebrandt.Waerbeke.2016, Troxel.Zhang.2018, Hikage.Yamada.2019, Hamana.Tanaka.2019, Asgari.Valentijn.2020, Amon.Weller.2021, Secco.To.2021, Dalal.Wang.2023, Li.Wang.2023}. Their findings, including hints of the $S_8$ tension and tight constraints on structure growth, have laid a strong foundation for the next generation of surveys.

Looking forward, Stage-IV surveys such as the Vera C. Rubin Observatory's Legacy Survey of Space and Time \citep[LSST;][]{Ivezic.Zhan.2019}, 
the European Space Agency’s Euclid mission \citep[Euclid;][]{Laureijs.Zucca.2011}, and the Nancy Grace Roman Space Telescope \citep[Roman;][]{Spergel.Zhao.2015} are poised to deliver unprecedented statistical power thanks to their vast survey areas, depth, and image quality. 
To fully exploit the information content of upcoming datasets, it is essential to develop and validate advanced analysis techniques using existing Stage-III data. In particular, HSC plays a unique role: it is not only one of the deepest wide-field imaging surveys to date, but also serves as a precursor to LSST. The methodologies refined on HSC data, including the control of systematics and the use of novel statistics, will directly inform the analysis strategies for LSST and other Stage-IV surveys such as Euclid.

So far, the two-point correlation function (2PCF) of the cosmic shear field has been the workhorse of weak lensing cosmology. However, as we aim for percent level precision in Stage-IV surveys, it is becoming increasingly evident that the 2PCF alone cannot capture the full statistical information—particularly the non-Gaussian features arising from nonlinear structure formation. 
To fully exploit the information encoded in the cosmic shear field, we must turn to higher-order statistics (HOSs). HOSs are sensitive to non-Gaussian signatures and provide access to complementary information beyond the 2PCF. They can help break parameter degeneracies, test for systematic errors, and improve the robustness and precision of cosmological constraints. Recent studies have demonstrated the potential of these statistics to enhance cosmological analyses; e.g. higher-order moments \cite{Chang.Collaboration.2018,Gatti.collaboration.2020, Gatti.Collaboration.2022,Peel.Baldi.2018, Petri.Kratochvil.2015,Porth.Smith.2021, Waerbeke.Velander.2013,Vicinanza.Er.2016, Vicinanza.Er.2018}, peak counts \cite{Ajani.Liu.2020, Dietrich.Hartlap.2010,Harnois-Deraps.Reischke.2021, Kacprzak.Collaboration.2016,Kratochvil.May.2010,Liu.May.2015,Martinet.Nakajima.2017,Peel.Baldi.2018,Shan.Wang.2017,Zurcher.Kacprzak.2023,Zurcher.Refregier.2020}, 
marked power spectra \cite{Cowell.Liu.2025}, one-point probability distributions \cite{Barthelemy.Gavazzi.2020,Boyle.Baldi.2021,Thiele.Smith.2020}, Minkowski functionals \cite{Grewal.Amon.2022,Kratochvil.Huffenberger.2012,Parroni.Scaramella.2019,Petri.Kratochvil.2015,Vicinanza.Tereno.2019}, Betti numbers \cite{Feldbrugge.Vegter.2019,Parroni.Scaramella.2020}, persistent homology \cite{Heydenreich.Martinet.2022,Heydenreich.Schneider.2022,Prat.Weller.2025}, scattering transform coefficients \cite{Cheng.Bruna.2020,Valogiannis.Dvorkin.2022}, wavelet phase harmonics \cite{Allys.Mallat.2020}, and kNN and CDF \cite{Anbajagane.Wiseman.2023,Banerjee.Abel.2022}, map-level inference \cite{Boruah.Hudson.2022,Porqueres.Lavaux.2021}, and machine-learning methods \cite{Fluri.Schneider.2019,Fluri.Hofmann.2018,Jeffrey.Lanusse.2020,Lu.Li.2023,Ribli.Csabai.2019}

Among various HOSs, the three-point correlation function (3PCF) represents a mathematically well-defined and well-studied extension of the 2PCF in real space \citep[See][for the early studies of the 3PCF]{Schneider.Lombardi.2002,Schneider.Lombardi.2003,Zaldarriaga.Scoccimarro.2003}. It is directly related to the bispectrum and has the advantage of being interpretable in configuration space, although it comes with increased computational and modeling complexity\citep{Takada.Jain.2004, Heydenreich.Schneider.2022}.
\citet{Porth.Schneider.2023} develop the new measurement algorithm of 3PCF from the galaxy shape catalog based on the same multipole decomposition, which reduces the computational time from the naive scaling $O(n^3)$ to $\mathcal{O}(n\log n)$.
Also, \citet{Sugiyama.Jarvis.2024} recently changed the situation by introducing the multipole decomposition method of shear 3PCF, which enables us to quickly compute the 3PCF from a given bispectrum model through fast Fourier transform \cite[see also][for the same approach applied to 3PCF of convergence field]{Arvizu.Collaboration.2024,Samario-Nava.Hidalgo}. 
Notably, \citet{Gomes.Weller.2025} developed a pipeline to enable cosmological inference from aperture-mass skewness, which is a physically motivated efficient compression of the 3PCF data vector, using DES-like mock data, and recently applied this methodology to actual DES Y3 data (Gomes et al. in prep.), performing a joint, 2PCF and aperture-mass skewness, cosmological analysis.

In this work, we apply this methodology to the public Year 3 (Y3) data from the HSC survey. We perform a joint analysis of the cosmic shear two-point and three-point correlation functions measured from the HSC-Y3 dataset, and derive constraints on cosmological parameters. This represents the first cosmological application of the three-point cosmic shear correlation from HSC data and provides a key step toward the full exploitation of higher-order statistics in current and future lensing surveys.

The structure of this paper is as follows. 
In Section~\ref{sec:data}, we describe the HSC-Y3 data and the mock catalogs used in this paper.
Section~\ref{sec:measurement} presents the measurement of the two- and three-point correlation functions from the data.
In Section~\ref{sec:model}, we detail the theoretical modeling of the correlation functions, including the multipole decomposition used for the 3PCF.
Section~\ref{sec:parameter-inference} outlines the parameter inference framework and likelihood analysis.
We present our main cosmological constraints and consistency checks together with the validation of our analysis pipeline
in Section~\ref{sec:result}.
Finally, we summarize our findings and discuss future prospects in Section~\ref{sec:conclusion}.

\section{Data}\label{sec:data}
\subsection{HSC-Y3 data}\label{sec:hscy3-data}
\subsubsection{HSC-Y3 shape catalog}\label{sec:shape-catalog}
In this paper, we use the HSC-Y3 shape catalog to probe the weak lensing signal. 
The shape catalog is based on the internal data release of S19A \cite{Aihara.Yuma.2018}, which was obtained in HSC-SSP from March 2014 to April 2019. 
\citet{Li.Yoshida.2022} generated the galaxy shape catalog based on the image data in S19A, which we call HSC-Y3 shape catalog. 
The above original shape catalog contains more than 35 million galaxies covering 433~${\rm deg}^2$. 
The shape catalog used for the science is obtained by applying the additional selection cuts on top of the original galaxies to have better quality of data as described in detail in \cite{Li.Wang.2023}.
The resulting shape catalog has 25 million galaxies covering 416~${\rm deg}^2$, and the effective number density 14.96~${\rm arcmin}^{-2}$. 
In this paper, we use the same shape catalog used in the real-space cosmic shear 2PCF analysis \cite{Li.Wang.2023} for a fair comparison.

\subsubsection{HSC-Y3 photometric redshift catalog}\label{sec:photo-z-catalog}
Here we introduce the photometric redshift (photo-$z$) estimates we use in this paper. We refer the readers to \citet{Nishizawa.Takata.2020} for more details. \citet{Nishizawa.Takata.2020} applied three different photo-$z$ code to estimate the photometric redshift for individual galaxies, \codett{dNNz}, \codett{DEMmP}, and \codett{MIZUKI}.
\citet{Rau.Takada.2022} developed the hierarchical Bayesian framework to combine all the results of the three photo-$z$ codes and clustering redshift obtained from CAMIRA LRG.
For the tomography analysis, we divided our galaxy samples into four redshift bins: $z\in(0.3, 0.6]$, $(0.6, 0.9]$, $(0.9,1.2]$, and $(1.2, 1.5]$.
Assignment of the galaxy samples to the redshift bins is based on the best estimate of the individual galaxies' redshift by \codett{dNNz}.
We found that some of the galaxy samples assigned to the first and second redshift bins exhibit secondary peak at high redshift ($z>3.0$) due to the degeneracy of the photo-$z$ code. This secondary peak is beyond the calibration redshift range of CAMIRA-RLGs ($z\leq1.2$), so we removed the galaxy samples which exhibit the secondary peak to avoid significant systematic bias in photo-$z$ estimate. We refer readers to \citet{Rau.Takada.2022} for more details and concrete criteria for selection of the galaxy with secondary peak. Fig.~2 of \citet{Li.Wang.2023} shows the stacked photo-$z$ distribution of the galaxy used for our measurements. In this paper, we mainly use the photo-$z$ estimate from combined photo-$z$ codes and CAMIRA-LRGs clustering redshift by \citet{Rau.Takada.2022}.

\subsubsection{HSC-Y3 star catalog}\label{sec:hscy3-star-catalog}
We use the star catalog to quantify the residual systematic bias due to the PSF modeling error on our correlation function which we introduce in the following sections. The star catalog we use in this paper is the same one as used in \citet{Li.Wang.2023}, where two different type of samples are defined, PSF stars and non-PSF stars. We use the PSF star sample to estimate the additive residual bias on the correlation functions. We refer further details of the star catalogs to \citet{Li.Wang.2023}.

\subsection{Mock shape catalogs}\label{sec:mock-catalogs}
We use the HSC-Y3 mock catalogs to estimate the covariance matrix of the data vector. These mock catalogs are the same mock catalogs used in the series of HSC-Y3 cosmology analyses \citep{Li.Wang.2023,Dalal.Wang.2023,More.Wang.2023,Sugiyama.Wang.2023,Miyatake.Wang.2023}. 
We refer readers to the papers for detailed information on these mock data, but here we briefly recap the basic information about the mock catalogs.
The mock catalogs are created by following the procedure of \citet{Shirasaki.Miyatake.2019} based on the full-sky simulation generated by \citet{Takahashi.Shiroyama.2017}. From each realization of the full-sky simulations, we cut out 13 HSC-Y3 footprints, resulting 1404 independent HSC-Y3 mock maps in total. For each HSC-Y3 mock map, we placed the galaxies at the same positions as in HSC-Y3 real data with shape-noise, and the cosmic shear is assigned based on the shear map to determine the galaxies' shapes.

\section{Measurements}\label{sec:measurement}
\subsection{Definition of the estimators}
In this section, we define the estimators of the 2PCF, 3PCF and the aperture-mass skewness measured in this paper.
From each of the shape catalogs introduced in Section~\ref{sec:data}, we measure the 2PCF and the aperture-mass skewness. The 2PCF between redshift bin $a$ and $b$ can be measured by
\begin{align}
    \hat{\xi}^{ab}_\pm(\theta) = \frac{\sum_{ij\in ab}w_iw_j[\gamma_{i,\rm t}\gamma_{j,\rm t}\pm\gamma_{i,\cross}\gamma_{j,\cross}]}{\sum_{ij\in ab}w_iw_j} \ ,
    \label{eq:2pcf-estimator}
\end{align}
where the summation runs over all the galaxy pairs of $i$ and $j$, from the redshift bin $a$ and $b$ respectively, with separation angle $\theta$. The shear projection, i.e. the tangential and cross components of the shear $\gamma_{t,\cross}$, are defined with respect to the line connecting the two galaxies. 
The 2PCF is measured for the separation angle $\theta$ from 0.276 arcmin to 333 arcmin and binned into equally spaced logarithmic 24 bins. 

For the measurement of aperture-mass skewness, we first measure the 3PCF of galaxy shears, which is namely the natural extension of 2PCF. Because the shear is a spin-2 quantity, which has two degree of freedom, the cosmic shear 3PCF has four modes. The choice of the modes is studied in \citet{Schneider.Lombardi.2002} and so-called natural components introduced in that paper is widely used, with which the modes are independent to each other and invariant up to a phase factor under the any exchange of the shear projections.  In this paper, we measure the natural component of the cosmic shear 3PCF, and especially we use the multipole-based estimator developed in \citet{Porth.Schneider.2023}. The estimator is defined as
\begin{align}
    \hat{\Gamma}^{\cross,abc}_\mu(\theta_1,\theta_2,\phi) = \frac{\sum_m \Upsilon_{\mu,m}^{abc}(\theta_1, \theta_2)e^{im\phi}}{\sum_m \mathcal{N}_m^{abc}(\theta_1,\theta_2)e^{im\phi}} \ ,
    \label{eq:3pcf:estimator}
\end{align}
where $\theta_1$ and $\theta_2$ are the two side lengths of the triangle connecting the three galaxies, $\phi$ is the opening angle between them, $\mu~(=0,1,2,3)$ is the index of natural components, and $m~(=-m_{\rm max},\cdots,m_{\rm max})$ is the multipole index.
In the multipole-based approach, the 3PCF is reconstructed by Fourier transforming the multipoles of galaxy triplets, $\Upsilon_{\mu, m}^{\cross,abc}$ and $\mathcal{N}_m^{abc}$. 
We refer readers to \citet[see Eqs. (21) and (23)-(25)]{Porth.Schneider.2023} for the explicit expressions for these galaxy triplets, but here we recap some important points. 
In the multipole-based approach, the computational cost to estimate the galaxy triplets scales as $\mathcal{O}(m_{\rm max}\times N_{\rm gal}\log N_{\rm gal})$ whereas the naive triplets counts scales as $N_{\rm gal}^3$. 
This huge gain in the computational efficiency is enabled by the decomposition of the triplet multipole expression into two identical position-dependent two-point functions, which requires the specific shear projection called $\cross$-projection introduced in \citet{Porth.Schneider.2023}. The superscript in Eq.~(\ref{eq:3pcf:estimator}) indicates the $\cross$-projection.

In this paper, we use the \code{TreeCorr}\footnote{\url{https://github.com/rmjarvis/TreeCorr}} implementation of 3PCF in Eq.~(\ref{eq:3pcf:estimator}). 
We bin the $\theta_1$ and $\theta_2$ into logarithmically spaced 20 bins from 1 arcmin to 80 arcmin, and use $m_{\rm max}=30$. 
When we measure the 3PCF for cross-redshift bins ($a=b\neq c$, $a=c\neq b$, $b=c\neq a$ or $a\neq b\neq c$), we do not force the ordering of the redshift bins to be the same as that of the triangle vertices, which can be accomplished by setting \code{ordering=False} in \code{TreeCorr}. 
For Fourier transformation in Eq.~(\ref{eq:3pcf:estimator}), we use linearly spaced $30$ points on $\phi$ from $5.24\times10^{-2}$ to $3.09$ radians.
However, we note that these points on $\phi$ do not mean the binning of the triangle, because in the multipole-based approach the $\phi$ dependence is reconstructed from the multipoles, and $\phi$ can be any real values in $[0,2\pi]$.

The aperture-mass skewness is linearly related to the cosmic shear 3PCF as
\begin{align}
\begin{split}
    \mapcube{}{,abc}(R)
    &= 
    \int_0^\infty\dd\theta_1
    \int_0^\infty\dd\theta_2
    \int_0^{2\pi}\dd\phi\\
    &\hspace{-3em}\times
    \frac{1}{4}\mathcal{R}\left[
    \sum_{\mu}\Gamma_\mu^{{\rm cen}, abc}
    (\theta_1,\theta_2,\phi)
    T_\mu\left(\frac{\theta_1}{R},\frac{\theta_2}{R},\phi\right)
    \right] \ ,
\label{eq:map3-integral}
\end{split}
\end{align}
where $\mathcal{R}$ indicates the real part of the quantity inside the bracket, and $T_\mu$ is the filtering function to transform the 3PCF to the aperture-mass skewness whose definition can be found in \citet{Jarvis.Jain.2003}. 
Note that the 3PCF here is defined with centroid projection, where the shear projection for each galaxy on the triangle vertex is defined with respect to the centroid of the triangle. 
The change of the shear projections introduces a phase factor for each natural component, and the factor can be found in \citep{Porth.Lee.2021}. 
To estimate the aperture-mass skewness from the measured 3PCF, we utilize the discretized version of the above equation, namely given by the Riemann sum:
\begin{align}
\begin{split}
    \hat{\mathcal{M}}_{\rm ap}^{3,abc}(R)
    &= \sum \Delta\!\ln\!\theta^2\theta^2\Delta\phi \\
    &\hspace{-3em}\times
    \frac{1}{4}\mathcal{R}\left[
    \sum_{\mu}\Gamma_\mu^{{\rm cen}, abc}
    (\theta_1,\theta_2,\phi)
    T_\mu\left(\frac{\theta_1}{R},\frac{\theta_2}{R},\phi\right)
    \right] \ ,
\label{eq:map3-riemann}
\end{split}
\end{align}
where the summation on the first line runs over the all the $\theta_1$ and $\theta_2$ bins, and $\phi$ points.
To perform the above summation including the shear projection change from $\cross$ to centroid projection, we use \code{toSAS} and \code{calculateMap} class methods of \code{TreeCorr} in this order. For the filter radius of the aperture-mass skewness $R$, we use logarithmically spaced 8 points from 1 arcmin to 50 arcmin.

As noted in \citet{Sugiyama.Jarvis.2024}, the notations of the triangle in \citet{Porth.Lee.2021} and \code{TreeCorr} are different and care is needed for the interpretation of the measured 3PCF.

Lastly, for the future reference, we also comment on the computational time of the 3PCF in Eq.~(\ref{eq:3pcf:estimator}) on the HSC-Y3 catalog. We used a computational server at Kavli IPMU and turned on \code{openmp} with 28 cores, with which a measurement of auto (cross) redshift bin 3PCF took 20 minutes (1 hour). The reason why the measurement for cross takes three times longer than that for auto is because we set \code{ordered=False} distinguishing the same triangle with different galaxy shear on vertices.

\subsection{Covariance}\label{sec:covariance}
\begin{figure*}
    \centering
    \includegraphics[width=\linewidth]{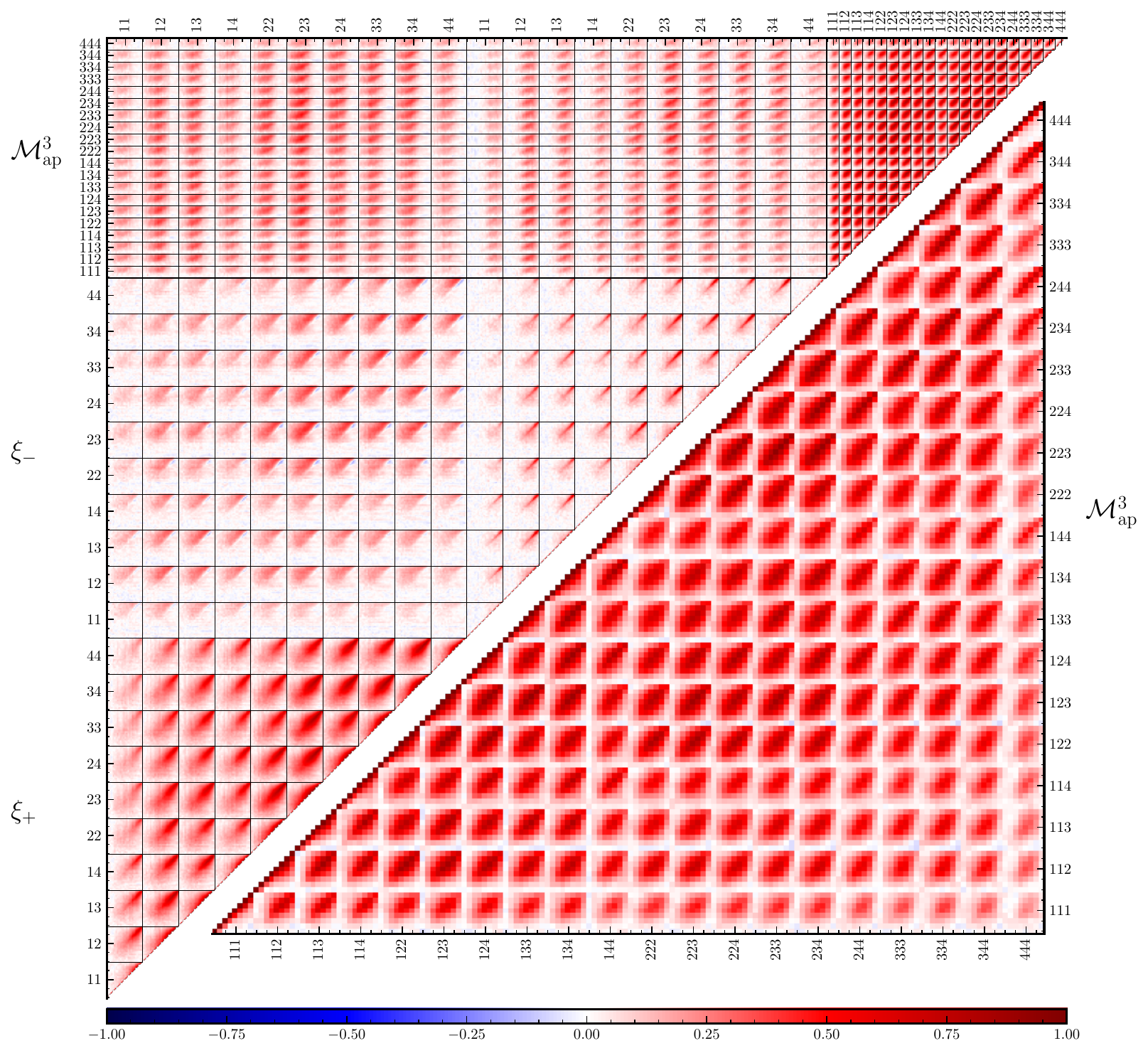}
    \caption{The correlation coefficient of the aperture-mass skewness estimated from the 1404 HSC-Y3 mock catalogs. The upper left triangle shows the correlation coefficients for the full data vector, which is a concatenation of the 2PCF and $\mapcube{}{}$. The lower right triangle shows a zoom-in of the correlation coefficients of $\mapcube{}{}$. The redshift bin combinations $ab$ and $abc$ are shown on the ticks for 2PCF and aperture-mass skewness respectively.
    2PCF has two modes ($\xi_\pm$), 10 redshift-bin combinations ($ab$) and 24 angular bins while the aperture-mass skewness has 20 redshift-bin combinations ($abc$) and 8 filter radii, yielding $2\times10\times24+20\times8=640$ elements for the full data vector. 
    The bottom color bar indicates the scale of correlation coefficients.}
    \label{fig:cov-map3}
\end{figure*}

We estimate the covariance matrix of the data vector by running the same measurement of the 2PCF and the aperture-mass skewness on the 1404 HSC mock catalogs introduced in Section~\ref{sec:mock-catalogs}.
In the mock catalog, the shell-structure is adopted to construct the mass map at each redshift point, where the mass map of each shell is constructed by projecting the 3D matter density field along the line-of-sight over the finite thickness of the shell. This introduces the mode-coupling and smoothing effect of the density field, and as a consequence the measured 2PCF from the mock catalog becomes typically smaller than the theoretical prediction at the scale of the interest. 

\citet{Li.Wang.2023} found that this shell thickness effect has no strong scale dependence or redshift dependence in the scale and redshift range of interest, and that the correction factor needed for measurement to match the theoretical prediction is $0.81$ on average. In this paper, we follow the same method to correct for the shell-thickness effect. For the 2PCF, we divide the measured signal by a 0.81 factor, and for the aperture-mass skewness, we divide by a $0.81^{1.5}$ factor as the order of the aperture-mass skewness is 1.5 larger than that of the 2PCF.

After measuring and correcting all the 2PCF and the aperture-mass skewness from each mock catalogs, we concatenate the measured signals into one data vector $\bm{d}^{(r)}$ for $r$-th mock catalog, and computed the sample covariance matrix as a estimate of the covariance matrix for the 2PCF and the aperture-mass skewness 
\begin{align}
\bm{C} = \frac{1}{1404}\sum_{r=1}^{1404} (\bm{d}^{(r)}-\bar{\bm{d}}) (\bm{d}^{(r)}-\bar{\bm{d}})^{\rm T} \ ,
\label{eq:cov}
\end{align}
where $\bar{\bm{d}}=\sum_{r=1}^{1404}\bm{d}^{(r)}/1404$ is the data-vector mean. Fig.~\ref{fig:cov-map3} shows the correlation coefficient of the estimated covariance matrix. The 2PCF has two modes ($\xi_\pm$), 10 redshift-bin combinations ($ab$) and 24 angular bins, and the aperture-mass skewness has 20 redshift-bin combinations ($abc$) and 8 filter radii, yielding $2\times10\times24+20\times8=640$ dimension for the full data vector.

\subsection{Signals from HSC-Y3 data}
\begin{figure*}
    \centering
    \includegraphics[width=\linewidth]{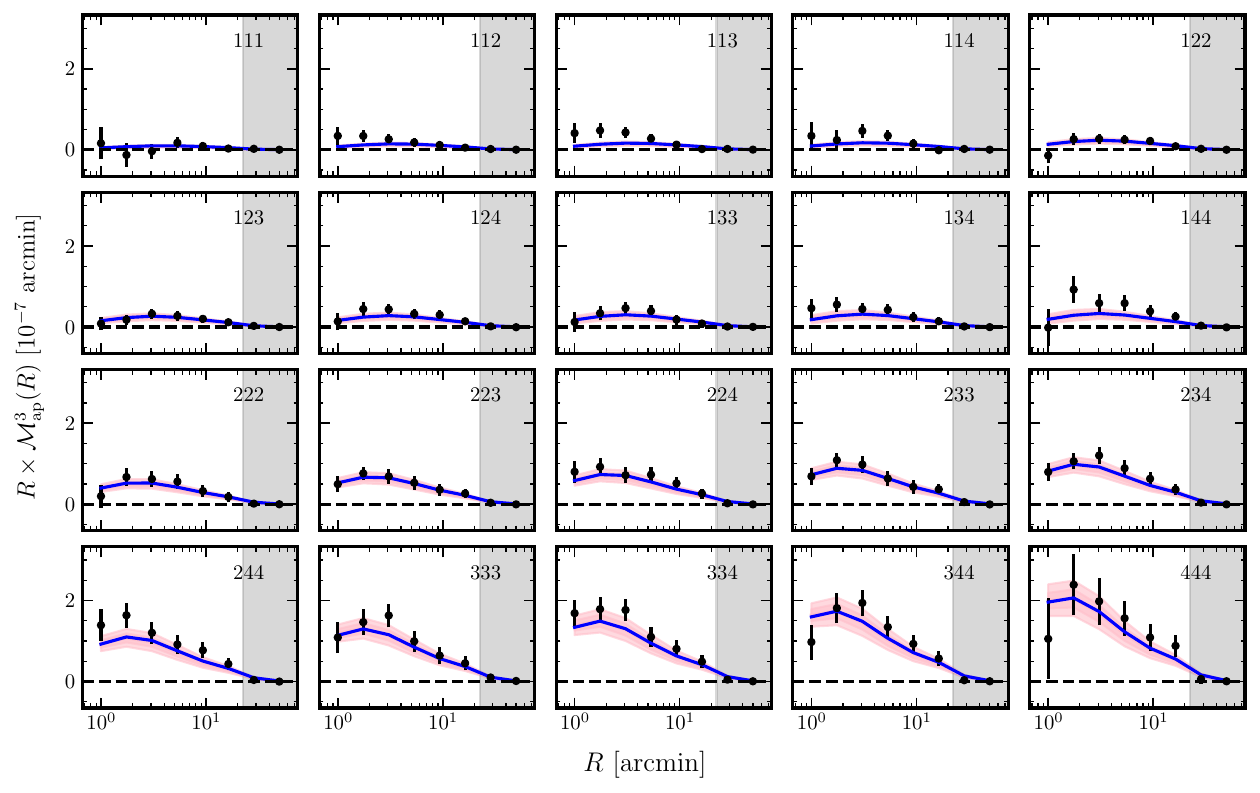}
    \caption{The aperture-mass skewness from the HSC-Y3 survey. Each panel shows the 
    $\mapcube{}{}$ as a function of filter radius $R$, and the combination of redshift bins $abc$ as indicated on the upper right of each panel. The black data points are the measurements with  error bars  estimated from  1404 mock catalogs. 
    The shaded region at large scales indicates the scale cut due to the observational systematics, which is related to the Field-of-View (FoV) scale of the HSC camera (1.5 degrees).
    The blue curve is the theoretical prediction with the best-fit model parameters obtained through the Bayesian parameter inference described in Section~\ref{sec:parameter-inference}. The pink regions are the 1- and 2-sigma credible intervals of the best fit. 
    }
    \label{fig:signal-map3}
\end{figure*}

\begin{figure}[t]
    \centering
    \includegraphics[width=\linewidth]{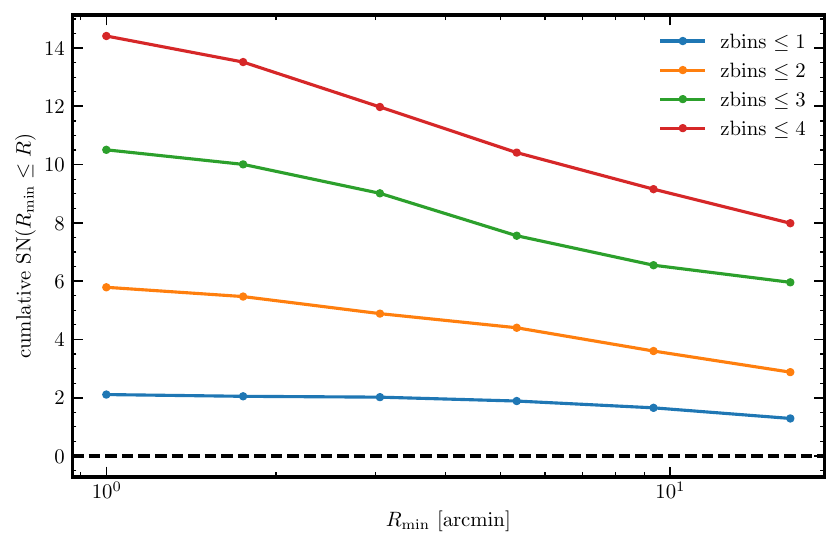}
    \caption{Cumulative signal-to-noise ratio (SNR) of the aperture-mass skewness  as a function of minimum filter radius. For each $R_{\rm min}$, the points show the SNR from all $R>R_{\rm min}$. From blue to red lines, we include successively higher redshift bins one by one, so the blue line shows the SNR for $abc\in\{111\}$, the orange line for $abc\in\{111, 112, 122, 222\}$, the green line for $abc\in\{111, 112, 113, 122, 123, 133, 222, 223, 233, 333\}$, and the red line for all 20 $abc$ combinations. Note that the data points at $R\geq 22.5$ arcmins indicated in the gray shaded region in Fig.~\ref{fig:signal-map3} are already discarded in this plot. }
    \label{fig:snr-map3}
\end{figure}

Fig.~\ref{fig:signal-map3} shows the measurement result of the aperture-mass skewness from HSC-Y3 data along with the error bar estimates from the 1404 HSC-Y3 mock catalogs. 
We first note that we discard the two data points at large scales in redshift-bin combination shown by the gray-shaded region in each panel. We have two reasons for this scale cut at large scale. The first reason is that we do not expect a large signal at this filter radius. Remembering the relation of the scales for 3PCF and aperture-mass skewness $\theta\sim 4R$ \cite[See Fig.~2 of][for the shape of the filter function $T_\mu$ relating $\theta$ and $R$]{Jarvis.Jain.2003} and the maximum measurement scale for 3PCF $\theta_{\rm max}=80$, the maximum filter scale where we expect signal is $R_{\rm max}\sim \theta_{\rm max}/4=20$ arcmins. Actually we observe that the signal rapidly drops at the scale of $R\geq 20$ arcmins. Another reason for this scale cut is the B-mode signal observed in \citet{Dalal.Wang.2023}, where they found that the B-mode signal becomes significant at the scale of Field-of-View (FoV) of HSC camera ($\theta_{\rm FoV}=1.5$ degrees). Again using the relation of scales of 3PCF and aperture-mass skewness, we obtain $R_{\rm FoV}=\theta_{\rm FoV}/4=22.5$. Although we did not find any significant B-mode signal in the aperture-mass skewness in Appendix~\ref{sec:null-test} because of the maximum measurement scale for 3PCF, we apply this scale cut to be conservative for the potentially remaining B-mode signal in the aperture-mass skewness and to be consistent with the scale cut on the 2PCF signal.

We can see that the signal is larger if the correlation includes a higher redshift bin, because of the longer line-of-sight for weak lensing.
We define the signal-to-noise ratio (SNR) of the signal as 
\begin{align}
    {\rm SNR}^2 = \sum_{ij} \bm{d}_i (\bm{C}^{-1})_{ij}\bm{d}_j \ ,
\end{align}
where $\bm{d}$ is the data vector, which is a concatenated array of the aperture-mass skewness signal measured from the HSC-Y3 shape catalog shown in Fig.~\ref{fig:signal-map3}. We found a total ${\rm SNR}=14.4$ in the aperture-mass skewness signal.
We also define the cumulative SNR as a function of the minimum filter radius $R_{\rm min}$ by restricting the summation only for the data vector elements satisfying $R_{\rm min}\leq R_i$.
Fig.~\ref{fig:snr-map3} shows the cumulative SNR ratio of the aperture-mass skewness as a function of minimum filter radius.
From the comparison of the different lines showing SNR including different redshift bins, we can see that the SNR receives contributions not only from highest redshift bin but from all redshift bins. 

Fig.~\ref{fig:cov-map3} shows the correlation coefficients of the data vectors.
We notice that the aperture-mass skewness has strong positive correlation between neighboring filter radii, which is because the aperture-mass skewness with neighboring filter radii receives the contribution of 3PCF at similar/same scales. We also notice that the aperture-mass skewnesses involving the common redshift bin(s) (e.g. 111 and 112) have stronger correlation than those involving no common redshift bins (e.g. 111 and 444). 
In the same figure, we also show the cross-covariance between 2PCF and aperture-mass skewness. We notice that the 2PCF and aperture-mass skewness are positively correlated on the measurement scales.

\section{Model}\label{sec:model}
\subsection{Weak lensing and intrinsic alignment}
In this section, we describe the model of the cosmic shear 2PCF and aperture-mass skewness. The cosmic shear 2PCF between redshift bins $a$ and $b$ can be expressed as Hankel transformation of the E and B mode angular power spectra:
\begin{align}
    \xi_\pm^{ab}(\theta) = \int_0^{\infty}\frac{\dd\ell \ell}{2\pi} J_{0/4}(\ell\theta)
    \left[C^{E;ab}(\ell)\pm C^{B;ab}(\ell)\right] \ ,
\end{align}
where $J_{0/4}(x)$ is the 0th/4th-order Bessel function of the first kind. 
The observed galaxy shapes are induced by the gravitational lensing causing the modification of the intrinsic galaxy shapes, $\gamma_{\rm obs}=\gamma_{\rm G}+\gamma_{I}$. 
The former correlates to the foreground matter density field, and is known to produce only E mode. 
The latter has spatial correlation, called intrinsic alignment (IA), because the intrinsic galaxy shapes are aligned to the underlying matter tidal field, and known to induce both of E and B mode in general. Therefore, the E and B mode angular power spectra of the observed galaxy shear can be expressed as
\begin{align}
    C^{E;ab} &= C^{E;ab}_{GG} + C^{E;ab}_{GI} + C^{E;ab}_{IG} + C^{E;ab}_{II} \ , \\
    C^{B;ab} &= C^{B;ab}_{II} \ .
\end{align}
There are two popular modeling choices for the intrinsic alignment, the tidal alignment and tidal torque (TATT) model and the nonlinear alignment (NLA) model. In TATT model, the galaxy IA has two contributions that are proportional to the matter tidal field and matter tidal torquing field, respectively. The NLA model is a subset model of the TATT model where only the tidal term computed with nonlinear matter power spectrum is included.
In this paper, we adopt the nonlinear alignment model (NLA), which only produces the E modes, motivated by the fact that no significant preference of NLA over TATT was found in the HSC-Y3 cosmic shear 2PCF analysis \cite{Li.Wang.2023}.

The angular power spectrum by the gravitational lensing effect is expressed as the line-of-sight integration of the matter density field, and under the flat sky and with the Limber approximation, its expression is given by
\begin{align}
    C_{GG}^{E;ab}(\ell) = \int_0^{\chi_{\rm H}}\dd\chi\frac{q_a(\chi)q_b(\chi)}{\chi^2}
    P_{\rm m}\left(\frac{\ell+1/2}{\chi}; z(\chi)\right) \ ,
\end{align}
where the $\chi$ is the comoving distance and $q_a(\chi)$ is the lensing efficiency for redshift bin defined by
\begin{align}
    q_a(\chi) = \frac{3\Omega_{\rm m}H_0^2}{2c^2}\frac{\chi}{a(\chi)}
    \int_\chi^{\chi_{\rm H}}\dd\chi' p_a(\chi') \frac{\chi'-\chi}{\chi'} \ .
\end{align}
Here $p_a(\chi')$ is the normalized redshift distribution of the source galaxies in the $a$-th redshift bin ($\int_0^{\chi_{\rm H}}\dd\chi p_a(\chi)=1$). For the matter power spectrum prediction, we use the \code{Halofit} model \cite{Takahashi.Oguri.2012}.

To model the intrinsic alignment, we need to model the GI and II part of the angular power spectra, but it can be included in a simple way for the case of NLA model. 
In NLA model, the galaxy intrinsic alignment is proportional to the nonlinear matter density field at the redshift of the source galaxy $\gamma_I\propto \delta_{\rm m}$ to which the lensing term is also proportional, and therefore the inclusion of NLA only requires the modification of the lensing efficiency kernel as:
\begin{align}
    q_a(\chi) \rightarrow q_a(\chi) + 
    f_{\rm IA}\left(z(\chi)\right)p_a(\chi)\frac{\dd z}{\dd\chi} \ .
    \label{eq:ia-kernel}
\end{align}
Here $f_{\rm IA}(z)$ describe the redshift dependence of the intrinsic alignment
\begin{align}
    f_{\rm IA}(z) = -A_{\rm IA}\left(\frac{1+z}{1+z_0}\right)^{\alpha_{\rm IA}}
    \frac{c_1\rho_{\rm crit}\Omega_{\rm m}}{D(z)} \ ,
\end{align}
where $z_0=0.62$ is a pivot redshift, $c_1\rho_{\rm crit}=0.0134$ is a conventional number, $D(z)$ is the linear growth factor, and $A_{\rm IA}$ and $\alpha_{\rm IA}$ are the parameters of the NLA model.

As the 2PCF, the cosmic shear 3PCF can also be expressed as the Fourier transformation of the convergence bispectrum:
\begin{align}
\begin{split}
    \Gamma^{\cross, abc}_\mu(\theta_1,\theta_2,\phi) 
    &= -\int\frac{\dd\bm\ell_1}{(2\pi)^2}\frac{\dd\bm\ell_2}{(2\pi)^2}e^{-i\bm\theta_1\cdot\bm\ell_2-i\bm\theta_2\cdot\bm\ell_2}\\
    &\hspace{-2em}
    \times b_{\kappa}^{abc}(\ell_1,\ell_2,\alpha)e^{2i\sum_i\beta_i} e^{i\varphi_\mu(\theta_1, \theta_2, \phi)} \ .
    \label{eq:3pcf-theory}
\end{split}
\end{align}
Here the convergence bispectrum is now parametrized by the two side lengths of the triangles in Fourier space $\ell_1$ and $\ell_2$, and the opening angle between them, $\alpha$. The angle $\beta_i$ is the polar angle of the Fourier mode $\bm\ell_i$ where $\bm\ell_3=-\bm\ell_1-\bm\ell_2$.
The last phase factor is solely determined by the triangle shape and independent from the absolute size of the triangle. The convergence bispectrum is expressed by the line-of-sight integration of the matter bispectrum
\begin{align}
\begin{split}
    b_\kappa^{abc}(\ell_1,\ell_2,\alpha) = 
    \int_0^{\chi_{\rm H}}\dd\chi\frac{q_a(\chi)q_b(\chi)q_c(\chi)}{\chi^4}\\
    \times B_{\rm m}
    \left(\frac{\ell_1}{\chi},\frac{\ell_2}{\chi},\frac{\ell_3}{\chi};z(\chi)\right) \ .
\end{split}
\end{align}
In order to include the consistent model of the intrinsic alignment as 2PCF, we consider the NLA model for 3PCF, which can be included just by replacing the lensing efficiency kernel by Eq.~(\ref{eq:ia-kernel}) in the above equation.
In this paper, we use the \code{BiHalofit} model for the matter bispectrum prediction \cite{Takahashi.Shirasaki.2019}. 

Performing the Fourier transformation in Eq.~(\ref{eq:3pcf-theory}) is computationally expensive because the integrand is oscillating and because the variable dependence of the integrand is inseparable. In this paper, instead of performing the Fourier transformation in Eq.~(\ref{eq:3pcf-theory}), we rely on the multipole based method developed in \citet{Sugiyama.Jarvis.2024}. In this approach, as have been done in the 3PCF measurement in Section~\ref{sec:measurement}, we decompose the opening angle dependence of the 3PCF and bispectrum into multipoles using the basis functions:
\begin{align}
    \Gamma_\mu^{\cross, abc}(\theta_1,\theta_2,\phi) 
    &= \frac{1}{2\pi}\sum_m \Gamma_{\mu, m}^{\cross, abc}(\theta_1,\theta_2) e^{im\phi} \ , \\
    b_\kappa^{abc}(\ell_1,\ell_2,\alpha)
    &= \sum_L b_{\kappa, L}^{abc}(\ell_1, \ell_2)P_{L}(\cos\alpha) \ ,
\end{align}
where $P_L(x)$ is the Legendre polynomial of degree $L$. The 3PCF multipoles can be expressed by the double Hankel transformation of bispectrum multipoles accounting for the multipole coupling functions, and this method is computationally efficient thanks to the use of 2DFFTLog for double Hankel transformation. 
We refer the readers to \citet{Sugiyama.Jarvis.2024} for more detail.

We use the \code{fastnc}\footnote{\url{https://github.com/git-sunao/fastnc}} to evaluate the multipole-based 3PCF. We use the same binning for $\theta_1$, $\theta_2$ and maximum multipole of 3PCF, $m_{\rm max}$, as measurement. Since the multipoles of 3PCF and bispectrum couple together, we use $L_{\rm max}=40$ for the maximum multipole of bispectrum. Once we obtained the 3PCF prediction on the same bins, we transform it to the aperture-mass skewness using Eq.~(\ref{eq:map3-riemann}). With this approach, we can account for the binning effect on the measured 3PCF consistently.

In this paper, we build an emulator of the aperture-mass skewness for efficient parameter inference by following the strategy in \citet{Gomes.Weller.2025}. We train the emulator based on the training data set generated by \code{fastnc} with varied model parameters. We leave the further detail of the emulator for Appendix~\ref{sec:map3-emulator}. 

\subsection{Observational systematics}
\subsubsection{Residual multiplicative bias}
The shape catalog introduced in Section~\ref{sec:hscy3-data} is calibrated using the realistic image simulations to accurately estimate the cosmic shear from the observed galaxy shape, while some uncertainty in the shear estimate can arise due to the assumption made in the calibration and the limited number of simulations \citep{Li.Yoshida.2022}. 
We model and marginalize over the uncertainty in the multiplicative bias on the shear estimate by introducing a nuisance parameter $\Delta m_a$ for $a$th redshift bin \citep{Amon.Weller.2021}. 
With this parametrization, the theoretical prediction of the 2PCF and aperture-mass skewness observables change as
\begin{align}
    \xi_\pm^{ab}(\theta)
    &\rightarrow (1+\Delta m_a)(1+\Delta m_b)\xi_\pm^{ab}(\theta) \ , \\
    \mathcal{M}_{\rm ap}^{3,abc}
    &\rightarrow (1+\Delta m_a)(1+\Delta m_b)(1+\Delta m_c) 
    \mathcal{M}_{\rm ap}^{3,abc} .
\end{align}
Because the multiplicative bias is calibrated and the accuracy is controlled at below 1\% level through the image simulations, we use the Gaussian prior with zero mean and a standard deviation of 0.01 when we perform parameter inference.

\subsubsection{Residual photo-$z$ bias}
The photometric redshift is estimated with the color band data of the HSC galaxy using the photo-$z$ code. Although the photo-$z$ code is calibrated with the spectroscopic data as described in \citet{Nishizawa.Takata.2020,Rau.Takada.2022}, the photometric redshifts for individual galaxies have residual bias due to the limited number of color band and the limited samples of spectroscopic data used to calibrate the photo-$z$ code. The residual photo-$z$ biases on individual galaxies propagate the uncertainty of the redshift distribution of each tomographic redshift bin.
In order to model and marginalize over the uncertainty of the redshift distribution for each tomographic bin, we introduce a nuisance parameter $\Delta\!z_a$, which shifts the mean galaxy redshift in $a$th tomography bin:
\begin{align}
    p_a(z) \rightarrow p_a(z+\Delta\!z_a) \ .
\end{align}
We use Gaussian priors on $\Delta\!z_a$ with zero mean and $\sim2\%$ standard deviations for the first and second tomography redshift bins. 
This choice is because the redshift range of the first two tomography bins contains reasonable number of spectroscopic and CAMIRA samples and because we believe the photo-$z$ to be well calibrated.
On the other hand, the last two tomography bins have only partial overlap with the redshift range of CAMIRA samples, and hence we use uninformative flat prior between -1 to 1 to allow the self-calibration of the residual photo-$z$ bias together with cosmological parameter in parameter inference.
This choice of the flat prior for the last two tomographic bins is motivated by the indications of the residual photo-$z$ bias at high redshift $z>1.2$, derived in the HSC-Y3 analyses \cite{Li.Wang.2023,Dalal.Wang.2023,Miyatake.Wang.2023,Sugiyama.Wang.2023}. 

\subsubsection{Residual PSF bias}
\begin{figure}[t]
    \centering
    \includegraphics[width=\linewidth]{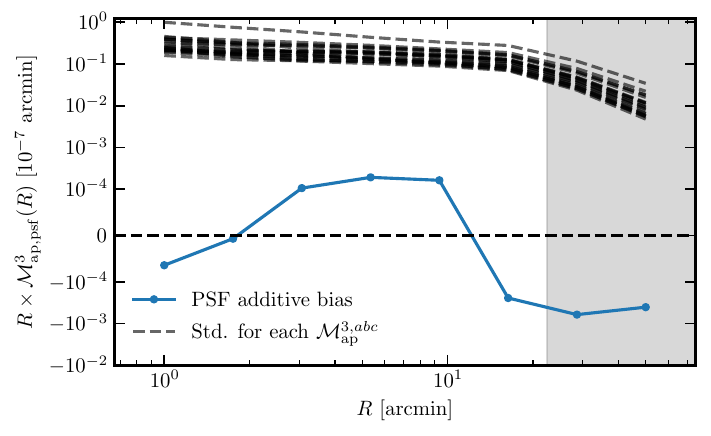}
    \caption{The PSF additive bias on the aperture-mass skewness (blue line) as predicted by the second term of Eq.~(\ref{eq:map3-psf}). 
    Note that all the PSF-terms in Eq.~(\ref{eq:map3-psf}) are  summed up in this figure. The black dashed lines show the standard deviations of the aperture-mass skewness in each redshift-bin combination $\mathcal{M}_{\rm ap}^{3,abc}$. The shaded region at large scale indicates the scale cut. The PSF additive bias for the aperture-mass skewness is two order of magnitude smaller than the standard deviation.}
    \label{fig:signal-map3-psf}
\end{figure}
The observed galaxy image is a pixelated convolution of the intrinsic galaxy profile and the Point Spread Function (PSF). Therefore, the PSF needs to be corrected to accurately recover the galaxy shape before PSF convolution.
The PSF model is calibrated using the stars, while the inaccuracy of the PSF model and interpolation of the models between star positions to galaxy position could lead to additive bias to the galaxy shape estimate: $\gamma\rightarrow\gamma+\gamma_{\rm psf}$. To model this residual PSF systematics, we follow the \citet{Zhang.More.2022}, where the PSF additive bias is models by four terms: second and fourth moments of PSF leakage and modeling errors,
\begin{align}
\begin{split}
    \gamma_{\rm psf} 
    &= 
    \alpha^{(2)}e_{\rm psf}^{(2)} + \beta^{(2)}\Delta e_{\rm psf}^{(2)} +
    \alpha^{(4)}e_{\rm psf}^{(4)} + \beta^{(4)}\Delta e_{\rm psf}^{(4)} \\
    &= \sum_{s\in {\rm psf}} p_sS_s \ , 
\end{split}
\end{align}
where we defined the residual-PSF-error parameter vector $\bm{p}=(\alpha^{(2)}, \beta^{(2)}, \alpha^{(4)}, \beta^{(4)})$, PSF moments vector $\bm{S}=(e_{\rm psf}^{(2)}, \Delta e_{\rm psf}^{(2)}, e_{\rm psf}^{(4)}, \Delta e_{\rm psf}^{(4)})$, and the summation in the above equation runs over the elements of this vector. The residual PSF error parameters are measured by cross correlating the galaxy shape and the PSF moments. Once we estimate the residual PSF parameters, we can compute the additive contamination of the PSF systematics to the cosmic shear signal,
\begin{align}
    \xi_+^{ab} \rightarrow \xi_+^{ab} + \sum_{st\in {\rm psf}}p_sp_t \hat\xi_{+,{\rm psf}}^{st} \ ,
\end{align}
where $\xi_{+,{\rm psf}}^{st}$ is 2PCF of the $s$th and $t$th elements of PSF moment vector, which are measured from the auto correlation of the PSF moments estimated from star catalog. In \cite{Zhang.More.2022}, it is shown that the PSF additive bias does not show significant redshift dependence and that the bias is negligible for $\xi_-$, so we neglect as well in this paper.
We extend this approach to 3PCF, namely the PSF additive bias is given by
\begin{align}
    \Gamma_\mu^{\cross, abc} \rightarrow \Gamma_\mu^{\cross,abc}+ 
    \sum_{stu \in {\rm psf}}p_sp_tp_u
    \hat\Gamma_{\mu, {\rm psf}}^{\cross, stu} \ .
\end{align}
or equivalently in terms of the aperture-mass skewness
\begin{align}
    \mathcal{M}_{\rm ap}^{3, abc} \rightarrow
    \mathcal{M}_{\rm ap}^{3, abc} + 
    \sum_{stu\in{\rm psf}} p_sp_tp_u \hat{\mathcal{M}}_{\rm ap, {\rm psf}}^{3, stu} \ .
    \label{eq:map3-psf}
\end{align}
Here $\hat\Gamma_{\mu, {\rm psf}}^{\cross, stu}$ and $\hat{\mathcal{M}}_{\rm ap, {\rm psf}}^{3, stu}$ are 3PCF and aperture-mass skewness of $s$th, $t$th and $u$th elements of PSF moment vector. 
We use the prior distribution for residual-PSF-error parameters that is estimated from the galaxy-PSF 2PCF in \citet{Zhang.More.2022}. 
Since the prior has degeneracy between four residual-PSF-error parameters, we introduce a new set of nuisance parameters $\{\alpha^{\prime(2)}, \beta^{\prime(2)}, \alpha^{\prime(4)}, \beta^{\prime(4)}\}$ by diagonalizing and standardizing the prior distribution obtained from galaxy-PSF 2PCF, for each of which we use Gaussian prior with zero mean and the unit standard deviation.

The predicted additive bias due to residual PSF error with the mean parameters on aperture-mass skewness (the second term in Eq.~(\ref{eq:map3-psf})) is shown in Fig.~\ref{fig:signal-map3-psf}. We found that the additive PSF bias on aperture-mass skewness is two order of magnitude smaller than the standard deviation of the signal, and therefore we decided to neglect the PSF additive bias in our model.

\section{Parameter Inference}\label{sec:parameter-inference}
\begin{table}
\caption{Model parameters and their priors used in the fiducial parameter inference in this paper. $\mathcal{U}(a,b)$ denotes a uniform prior between $a$ and $b$, and $\mathcal{N}(\mu,\sigma)$ denotes a normal distribution with mean $\mu$ and width $\sigma$. The equation-of-state (EoS) parameter of dark energy $w$ is fixed to $w=-1$ for $\Lambda$CDM analysis.}
\label{tab:parameters}
\setlength{\tabcolsep}{15pt}
\begin{ruledtabular}
\begin{center}
\begin{tabular}{ll}
Parameter & Prior \\ \hline
\multicolumn{2}{l}{\hspace{-1em}\bf Cosmological parameters}\\
$\Omega_{\rm m}$        & ${\cal U}(0.1,0.7)$\\
$S_8$                   & ${\cal U}(0.63, 0.92)$\\
$n_{\rm s}$             & ${\cal U}(0.87,1.07)$\\
$h_0$                   & ${\cal U}(0.62,0.80)$\\
$\omega_{\rm b}$        & ${\cal U}(0.02, 0.025)$\\
$w$                     & ${\cal U}(-2.0, -0.5)$\\
\multicolumn{2}{l}{\hspace{-1em}\bf Intrinsic Alignment parameters}\\
$A_{\rm IA}$            & ${\cal U}(-6, 6)$ \\
$\alpha_{\rm IA}$       & ${\cal U}(-6, 6)$ \\\hline
\multicolumn{2}{l}{\hspace{-1em}\bf Photo-$z$ systematics}\\
$\Delta\!z_1$           & ${\cal N}(0, 0.024)$\\
$\Delta\!z_2$           & ${\cal N}(0, 0.022)$\\
$\Delta\!z_3$           & ${\cal U}(-1,1)$  \\
$\Delta\!z_4$           & ${\cal U}(-1,1)$  \\
\multicolumn{2}{l}{\hspace{-1em}\bf Shear calibration biases}\\
$\Delta\!m_1$           & ${\cal N}(0.0,0.01)$ \\
$\Delta\!m_2$           & ${\cal N}(0.0,0.01)$ \\
$\Delta\!m_3$           & ${\cal N}(0.0,0.01)$ \\
$\Delta\!m_4$           & ${\cal N}(0.0,0.01)$ \\
\multicolumn{2}{l}{\hspace{-1em}\bf PSF residuals}\\
$\alpha^{\prime(2)}$     & ${\cal N}(0,1)$ \\
$\beta ^{\prime(2)}$     & ${\cal N}(0,1)$ \\
$\alpha^{\prime(4)}$     & ${\cal N}(0,1)$ \\
$\beta ^{\prime(4)}$     & ${\cal N}(0,1)$ \\
\end{tabular}\end{center}
\end{ruledtabular}
\end{table}

Our parameter inference is based on the Bayesian theory, where the posterior distribution of the model parameters $\bm p$ for a given data $\bm d$ is given by the product of the likelihood of data and prior:
\begin{align}
    \mathcal{P}(\bm{p}|\bm{d}) \propto \mathcal{L}(\bm{d}|\bm{p}) \Pi(\bm{p}) \ ,
\end{align}
where $\mathcal{L}$ and $\Pi$ is the likelihood and prior.
We adopt a Gaussian likelihood for the data vector:
\begin{align}
    \ln\mathcal{L}(\bm d|\bm p) = 
    -\frac{1}{2}
    [\bm d - \bm t(\bm p)]^{\rm T}
    \bm{C}^{-1}
    [\bm d - \bm t(\bm p)] + {\rm const} .
    \label{eq:likelihood}
\end{align}
Here $\bm t(\bm p)$ is the theoretical prediction by model parameter $\bm p$ for the data vector $\bm d$.
In our fiducial analysis, the data vector is the concatenated data vector of the cosmic shear 2PCF and the aperture-mass skewness. 
The covariance matrix $\bm{C}$ is estimated from the 1404 measurements of the data vectors from the HSC-Y3 mock catalogs as described in Section~\ref{sec:covariance}.

In Table~\ref{tab:parameters}, we summarize the model parameters and their priors used in this paper. 
We note that most of this choice of the model parameters and priors are the same as in \citet{Li.Wang.2023,Dalal.Wang.2023}. 
However, we recap some of the basic and important points of this choice below for this paper to be self-contained.
We categorized the model parameters into two categories: the physical parameters and the nuisance parameters for the observational systematics. 
The physical parameters have five (six) cosmological parameters in $\Lambda$CDM ($w$CDM) model, 
and two intrinsic alignment parameters. 
We choose $S_8$ as the free parameter of the large-scale structure amplitude, because the parameter support range of the emulator is bounded by this parameter. 
Although the baryonic feedback model was adopted in \citet{Li.Wang.2023,Dalal.Wang.2023}, we do not include them in our model because we do not have a reliable baryonic feedback model on three-point statistics. In this paper, we instead adopt a scale cut at small scale on the data vector, and validate the choice using the mock data that incorporates the typical baryonic feedback on the 2PCF and aperture-mass skewness.
The nuisance parameters include four mean photo-$z$ bias parameters, four shear multiplicative bias parameters, and four PSF residual parameters.
As discussed in the HSC-Y3 papers \citep{Li.Wang.2023,Dalal.Wang.2023,Sugiyama.Wang.2023,Miyatake.Wang.2023}, we found that photometric redshifts for high redshift galaxies ($z\gtrsim1.2$) are less calibrated compared to the lower redshift due to limitation of the calibrating spectroscopic samples. 
Therefore, we allow the mean redshifts to freely vary for third and forth redshift bins by adopting the uninformative prior on $\Delta\!z_3$ and $\Delta\!z_4$.

We implement our analysis pipeline within the framework of \code{CosmoSIS}\cite{Zuntz.Kowalkowski.2014} that provides easy interfaces for parameter inference. We use the \code{cosmosis-standard-library} for the standard calculation of cosmology-related statistics and the collection of modules developed for HSC-Y3 real-space 2PCF analysis \footnote{\url{https://idark.ipmu.jp/~xiangchong.li/cosmic_shear_real_public/}}, both of which are publicly available. We additionally build \code{cosmosis-3pt-library} for the modeling of third-order shear statistics which is based on the codes developed in \citet{Gomes.Weller.2025} with some extensions dedicated for HSC-Y3 data \footnote{\url{https://github.com/git-sunao/hscy3-3pcf}}.
We Monte-Carlo sample the posterior distribution using \code{MultiNest} through \code{CosmoSIS} interface, and estimate the marginalized posterior distribution by \code{getdist}.

In this paper, we did not perform any blind analysis.

\subsection{Data compression}
For now, the data vector has 640 dimensions, which is quite high compared to the number of simulations (mock catalogs in our case) used for the covariance estimate. It is known that the estimate of the entire covariance matrix can be noisy and the noise can propagate to inverse covariance computation and the estimation of parameter posterior. Especially when the number of the simulations used for the covariance estimate is small compared to the dimension of the data vector and the number of parameters, the noise propagation can be severe to lead to the unreliable size of the covariance and the credible parameter region, which is known as the Anderson-Hartlap effect\cite{Anderson.Anderson.2003,Hartlap.Schneider.2007} and the Dodelson-Schneider effect \cite{Dodelson.Schneider.2013}. These papers suggest the correction factors that should be multiplied on the inverse of the estimated covariance matrix from the finite number of simulations. 
However, the covariance matrix estimate itself is a random variable, as the simulations have their own randomness inherently, but these correction factors is the correction on the average bias and can be not enough/too much depending on the realization of the simulations.

To overcome this difficulty of the dimensionality, we perform the data compression. There are various data-compression methods, e.g. principal component analysis (PCA), canonical correlation analysis (CCA), machine-learning based methods, and etc.,  each of which maximizes the different objectives through different methods \cite[See e.g.][for comprehensive study of the data compression in the context of weak lensing statistics]{Park.Jain.2024}. In this paper, we use so-called MOPED (Massively Optimised Parameter Estimation and Data) compression \cite{Heavens.Lahav.1999}. MOPED is a linear compression algorithm that reduces the original data vector into the smaller data vector with the same number of dimensions as the model parameters.
\begin{align}
    \bm{d} \rightarrow \bm{B}\cdot\bm{d} \ ,
\end{align}
where $\bm{B}$ is the compression matrix. In MOPED, the compression matrix is formulated to preserve Fisher information of the likelihood (therefore it is optimal in terms of Fisher information), so the compression utilizes the derivatives of the model prediction with respect to the model parameters as follows
\begin{align}
    \bm{B}_{i} = \bm{C}^{-1} \frac{\partial \bm{t}}{\partial p_i}\label{eq:moped-b} \ ,
\end{align}
which, by multiplying to the original data vector, reduces it to one single data point that is sensitive to the $i$-th model parameter, corresponding to the $i$-th element of the compressed data vector. Although the algorithm is followed by Gram–Schmidt process to make the compressed data vector orthonormal, the above equation is the core part of the MOPED compression to preserve Fisher information.

As can be seen in Eq.~(\ref{eq:moped-b}), the compression matrix includes the inverse covariance matrix of the original data vector, which can be unreliable when the data dimension is large. The authors of their following paper \cite{Heavens.Vianello.2017} recommend that we firstly run many simulations so that we can obtain a reliable estimate of the inverse covariance and the compression matrix, and then we perform the data compression to obtain the reliable covariance matrix for the compressed data vector without running many simulations. However, this approach is not affordable in our case as the number of the mock catalogs are already fixed to 1404, and we cannot obtain the compression matrix with an unreliable noisy covariance matrix estimate. 

To overcome this problem, we instead use a ``powered'' covariance matrix for the estimate of the compression matrix $\bm{B}$; $C^p_{ij}\equiv {\rm sign}(r^{cc}_{ij})\sigma_i\sigma_j|r^{cc}_{ij}|^f$. Here $\sigma_i$ is the standard deviation of the $i$-th data element, $r^{cc}_{ij}$ is the correlation coefficients, and $f$ is a parameter that we choose to avoid noise propagation. By using a large $f$, we can reduce the contribution of the off-diagonal elements and therefore reduce the severe noise propagation, while the structure of covariance is lost. Setting $f\rightarrow\infty$ is equivalent to removing the off diagonal elements and using the diagonal element of the covariance matrix. Conversely, by using a small $f\sim 1$, we keep the structure of the covariance, while the noise propagation gets larger. We use $f=[(1404-1)/(1404-\dim(\bm{d}))]^{1/2}$. 
The use of the powered covariance matrix introduces its own Dodelson-Schneider factor on the credible region of the parameters, which we estimate through the simulations of the covariance estimate where the samples are drawn from the covariance matrix in Eq.~(\ref{eq:cov}).
We found the Dodelson-Schneider factors are 9\%, 8\% and 14\% for 2PCF-only, aperture-mass-skewness-only, and joint analysis, respectively. The advantage of using the powered covariance is more investigated in another paper (Sugiyama et al. in prep.).

In the joint or 2PCF-only analysis, we compress the full or 2PCF data vector into a compressed data vector of dimension 19, while in the mass-skewness-only analysis, we compress the aperture-mass skewness data vector into size 15 because we do not include residual PSF model in the aperture-mass skewness. When we extend the model to $w$CDM, the dimension of the compressed data vector increases by one corresponding to the dark energy EoS parameter. Hereafter, we use the compressed data vector for the parameter inference, and do not include the Anderson-Hartlap factor in the likelihood since they are negligible after compression.

\section{Result}\label{sec:result}
\subsection{HSC-Y3 result}\label{sec:hsc-y3-result}
\begin{figure*}[t]
    \centering
    \includegraphics[width=\linewidth]{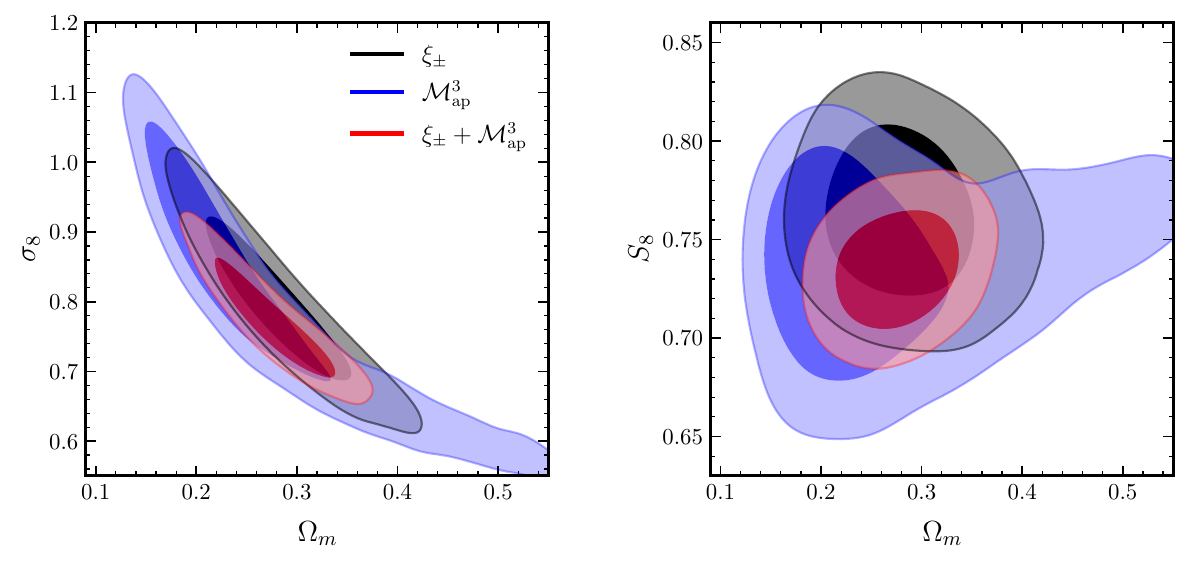}
    \includegraphics[width=\linewidth]{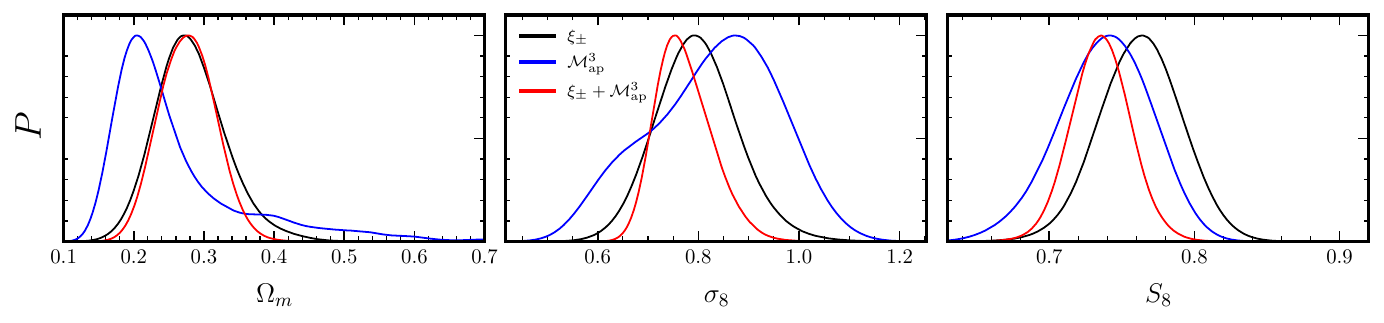}
    \caption{The cosmological parameter constraint from the joint analysis of the  2PCF ($\xi_\pm$) and aperture-mass skewness ($\mapcube{}{}$) measured from Subaru HSC Year 3 
    data in $\Lambda$CDM model. The upper panels are the marginalized 2D posterior distributions and the lower ones are the marginalized 1D posterior distributions. In each panel, the black, blue and red contours are the posteriors obtained from the 2PCF-only, $\mapcube{}{}$ only, and the joint analyses, respectively. The elongation of $(\Omega_{\rm m}, S_8)$-posterior in the upper right panel is because its preference for relatively low $\Omega_{\rm m}$ values, as demonstrated in the analysis on mock data (See Fig~\ref{fig:mock-2d-oms8-om02}).}
    \label{fig:data-2d-1d-omsig8-oms8}
\end{figure*}

\begin{table}
\caption{The result of the goodness-of-fit test of three fiducial analyses. In each row, the observed chi-squared value $\chi^2$, the effective degree of freedom of the data accounting for the model parameter freedom which is estimated with the Gaussian linear model \cite{Raveri.Hu.2019}, and the $p$ value are shown for each fiducial analysis. Note that $\chi^2$ is computed for the compressed data vector, so it is smaller than is typically expected from the dimension of the original data vector.}
\label{tab:goodness-of-fit}
\setlength{\tabcolsep}{15pt}
\begin{ruledtabular}
\begin{center}
\begin{tabular}{llll}
analysis & $\chi^2$ & $n_{\rm eff}$ & $p$ \\ \hline
2PCF-only          & 9.70 & 13.4 & 0.75 \\
skewness-only & 6.35 & 15.6 & 0.88 \\
joint              & 12.9 & 13.0 & 0.46 \\
\end{tabular}
\end{center}
\end{ruledtabular}
\end{table}

\begin{figure*}[t]
    \centering
    \includegraphics[width=\linewidth]{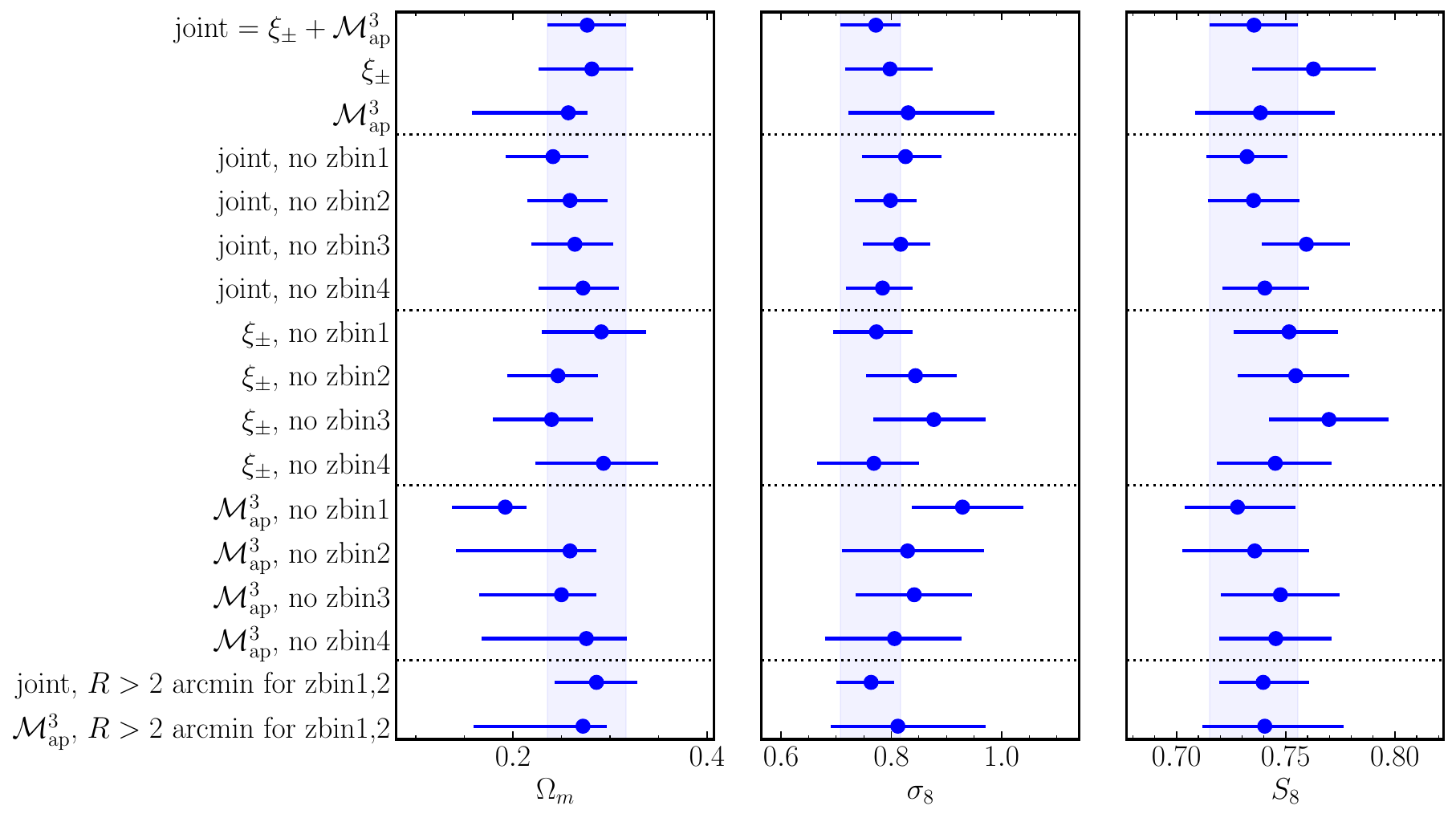}
    \caption{The result of  internal consistency tests, which shows the  parameter constraints on $\Omega_{\rm m}$, $\sigma_8$, and $S_8$ from different sub-samples or sub-probes compared to the fiducial joint analysis. In this  plot, the mean of the posterior is indicated by the point, and the error-bar indicate the $1\sigma$  interval. The top section shows the results of the fiducial joint analysis ($\xi_\pm+\mathcal{M}_{\rm ap}^3$), 2PCF-only ($\xi_\pm$) and aperture-mass-skewness-only ($\mathcal{M}_{\rm ap}^3$) analyses. The next three sections show the variants of these fiducial joint analysis where one redshift bin is removed from the analysis in each row. In these analyses, we adopt the photo-$z$ prior on $\Delta z_{3,4}$ that is obtained from the fiducial 2PCF-only analyses as done in \citet{Li.Wang.2023,Dalal.Wang.2023}. The last section shows the results of the analyses where we remove two small scale ($R\leq 2$ arcmins) data points of the aperture-mass skewness  for the two lowest redshift bins. This is intended to test the potential unknown systematic bias due to the baryonic effects on small physical scales
    (since lower redshifts probe smaller physical scales for a fixed angular scale). We do not find significant shifts relative to the fiducial analyses, which provides validation of our analysis.}
    \label{fig:data-whisker}
\end{figure*}

\begin{figure}[t]
    \centering
    \includegraphics[width=\linewidth]{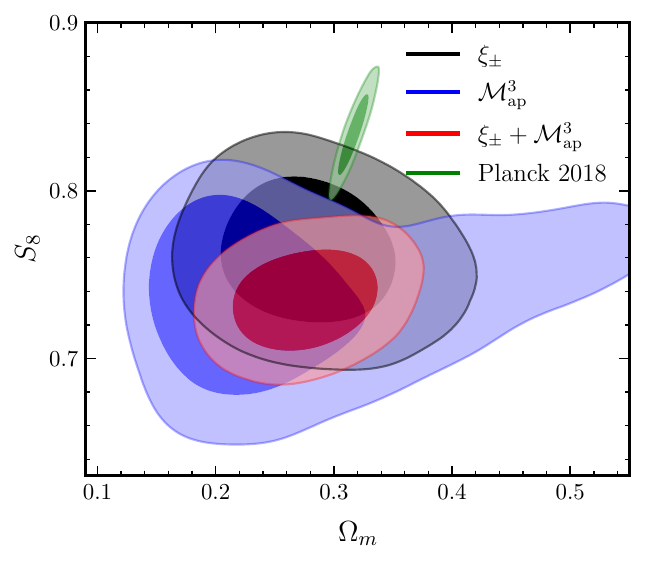}
    \caption{A comparison of the cosmological constraints from the HSC-Y3 2PCF+aperture-mass skewness joint analysis to the CMB constraint from Planck 2018 \cite{Collaboration.Zonca.2020}.}
    \label{fig:data-2d-comparison-planck}
\end{figure}

\begin{figure}[t]
    \centering
    \includegraphics[width=\linewidth]{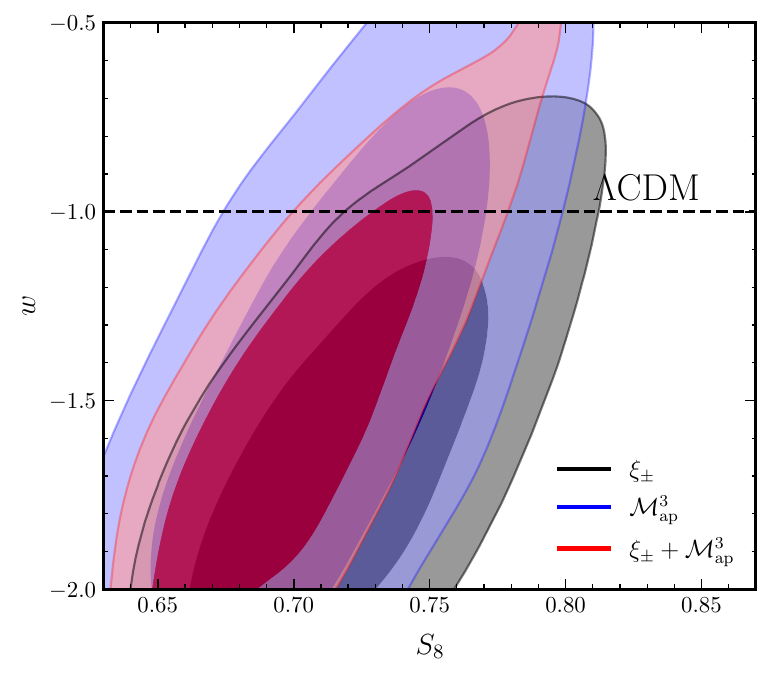}
    \caption{The cosmological parameter constraint in $w$CDM model. The same colors are used as in Fig.~\ref{fig:data-2d-1d-omsig8-oms8} for different probes. The slice at $w=-1.0$ indicated by the black dashed line corresponds to $\Lambda$CDM model. The EoS parameter $w$ is sampled within the prior range of $[-2.0, -0.5]$, and hence the posterior is bounded at lower edge $w=-2.0$ by this choice. Therefore we should {\it not} take the face value of the posterior mean as the result of the parameter constraint.}
    \label{fig:data-2d-wcdm}
\end{figure}

\begin{figure}[t]
    \centering
    \includegraphics[width=\linewidth]{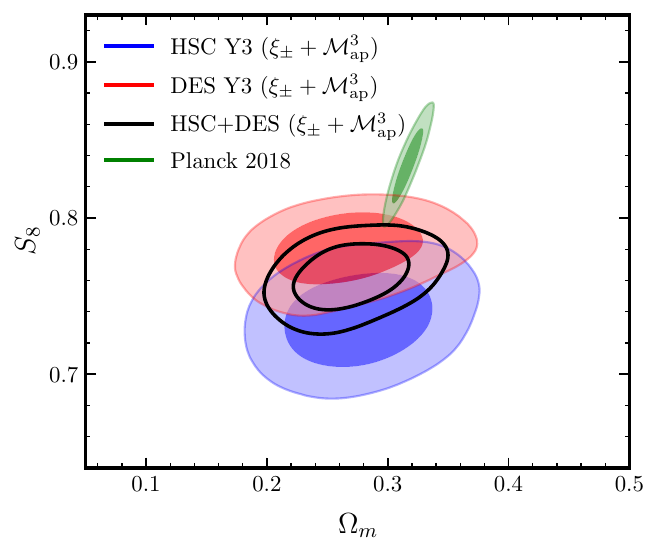}
    \caption{The joint posterior of 2+aperture-mass skewness from HSC-Y3 and DES-Y3 data. Here we assume the independence of the HSC-Y3 and DES-Y3 data vectors because the footprint overlap is not large, and the difference in the depth. The HSC-Y3 (blue) and DES-Y3 (red) posteriors are combined at the level of posterior in $(\Omega_{\rm m}, S_8)$ to obtain the joint posterior (black).}
    \label{fig:data-2d-hsc-x-des}
\end{figure}

In this section, we present the result of parameter inference from the HSC-Y3 data. Fig.~\ref{fig:data-2d-1d-omsig8-oms8} shows the 1D/2D marginalized posterior distributions for the main cosmological parameters $\Omega_{\rm m}$, $\sigma_8$, and $S_8$. First of all, we did not find any significant discrepancy in posteriors between the 2PCF-only and the aperture-mass-skewness-only analysis, which allow the joint analysis of them. The cosmological constraints we obtained from the joint analysis can be summarized as 
\begin{align}
\text{joint = $\xi_\pm$+$\mathcal{M}_{\rm ap}^3$:}
\begin{cases}
    \Omega_m &= 0.277\pm 0.040\\
    \sigma_8 &= 0.772^{+0.045}_{-0.064}\\
    S_8 &= 0.736\pm 0.020
\end{cases} \ ,
\end{align}
achieving a $3$\% constraint on the primary cosmological parameter $S_8$. Here, we report the mean of the posterior together with the $1\sigma$ highest density interval.
This can be compared to the results of the 2PCF-only analysis
\begin{align}
\text{$\xi_\pm$:}
\begin{cases}
    \Omega_m &= 0.281^{+0.043}_{-0.055}\\
    \sigma_8 &= 0.798\pm 0.082\\
    S_8 &= 0.763\pm 0.028
\end{cases} \ ,
\end{align}
and the aperture-mass-skewness-only analysis
\begin{align}
\text{$\mathcal{M}_{\rm ap}^3$:}
\begin{cases}
    \Omega_m &= 0.257^{+0.020}_{-0.10}\\
    \sigma_8 &= 0.83^{+0.16}_{-0.11}\\
    S_8 &= 0.738^{+0.034}_{-0.030}
\end{cases} \ .
\end{align}
Note that the cosmological constraint from the 2PCF-only analysis is slightly different from the result in \citet{Li.Yoshida.2022} because of the differences in the IA modeling and parametrization of the amplitude of power spectrum,  while the difference is less than $0.3\sigma$.

We quantify the Figure-of-Merit (FoM) of the parameter constraining power 
\begin{align}
    {\rm FoM}_{\Omega_{\rm m},S_8} = [\det C_{\Omega_{\rm m}, S_8}]^{-1/2} \ ,
\end{align}
where $C_{\Omega_{\rm m}, S_8}$ is the covariance matrix of the posterior samples for $\Omega_{\rm m}$ and $S_8$. 
The joint analysis yielded ${\rm FoM}_{\Omega_{\rm m},S_8}^{\rm joint}=1268$, achieving 80\% gain compared to the 2PCF-only analysis with ${\rm FoM}_{\Omega_{\rm m},S_8}^{\xi_\pm}=703$. This significant gain in FoM of the joint analysis is thanks to the different degeneracy directions in the 2PCF-only and the aperture-mass-skewness-only analyses. As seen in the upper left panel, the 2PCF is most sensitive to the combination of $\sigma_8\Omega_{\rm m}^{0.5}$, deriving the conventional $S_8$ definition, while the aperture-mass skewness is more sensitive to $\sigma_8\Omega_{\rm m}^{\alpha}$ with $\alpha<0.5$, and therefore the joint constraint breaks the degeneracies in $(\Omega_{\rm m},\sigma_8)$ or equivalently in $(\Omega_{\rm m}, S_8)$ \cite[See][for the scaling of third-order weak-lens statistics to the cosmological parameters]{Bernardeau.Mellier.1996}.

We also quantified the goodness-of-fit of the best fit model to the data in the three analyses. To assess the goodness-of-fit, we followed the way of \citet{Raveri.Hu.2019}; we use the Gaussian linear model (GLM) to obtain the effective degree of freedom. Table~\ref{tab:goodness-of-fit} summarizes the result of the goodness-of-fit test. In all individual analyses, we found the model fits reasonably well to the data.

We note that the posterior distribution from the aperture-mass-skewness-only analysis has a skewed shape in $(\Omega_{\rm m}, S_8)$ plane. We found that this is due to the slight preference in relatively low-$\Omega_{\rm m}$ due to the statistical scatter of the aperture-mass skewness relative to 2PCF. In Appendix~\ref{sec:low-om-mock}, we explicitly demonstrated that the posterior distribution is skewed as observed in Fig.~\ref{fig:data-2d-1d-omsig8-oms8}.

We did not find any significant signal of intrinsic alignment parameter from the joint analysis. Also we did not find significant improvement on the residual photo-$z$ bias parameter from the joint analysis. We leave the comprehensive contour plot for these parameters in Fig.~\ref{fig:data-triangle} in Appendix~\ref{sec:large-triangle-contour}.

To assess the internal consistency of the data vector measured from the HSC-Y3 data, we remove different parts of the data vector and check the variations relative to the fiducial result. Fig.~\ref{fig:data-whisker} shows the result of the internal consistency test. To test the systematics across the redshift bin, we removed one redshift bin from the analysis. 
In this analysis, in order to avoid a significant degradation of the constraining power, we adopted Gaussian priors on them $\Delta z_3\sim \mathcal{N}(-0.115,0.55)$ and $\Delta z_4\sim \mathcal{N}(-0.192,0.088)$ for joint, 2PCF-only, and aperture-mass-skewness-only analyses, following \citet{Li.Wang.2023,Dalal.Wang.2023}.
The result of this analysis removing one redshift bin is shown in the second, third and fourth sections of the figure for the joint, 2PCF-only, and the aperture-mass-skewness-only analyses, respectively. We did not find any significant shift larger than $1\sigma$. 
To test the baryonic feedback effect at small scale, we removed the smallest scale from the aperture-mass skewness data vector by restricting the filter radii $R\geq2$ arcmins in the two lowest redshift bin (zbin1 and zbin2). This is because the baryonic effect is more important at small scale in physical dimension $k\gtrsim1 h{\rm Mpc}^{-1}$, and because the lower redshift probes a smaller physical scale for a fixed angular scale, $\ell\sim k\chi$. The result is shown in the fifth section of the figure, and again we did not find any significant shift, which indicates that the data do not show any significant baryonic effect on the aperture-mass skewness. Based on these results in this paragraph, we conclude the internal consistency of our data vector.

Fig.~\ref{fig:data-2d-comparison-planck} compares the fiducial cosmological constraint obtained in this paper to the CMB result from Planck 2018\cite{Collaboration.Zonca.2020} in $\Lambda$CDM cosmology. 
Following the KDE method developed in \cite{Raveri.Doux.2021}, we find that the tension between the fiducial result and the Planck-2018 CMB is $3.2\pm0.1\sigma$ in the 2D space.

We extend the model to $w$CDM by freeing the EoS parameter of dark energy, $w$. Fig.~\ref{fig:data-2d-wcdm} shows the result. The degeneracy in $S_8$ and $w$ is not entangled because both the 2PCF-only and the aperture-mass-skewness-only analysis shows the same degeneracy direction in this parameter space, while the constraint is tighten in $S_8$ direction by the joint analysis. We conclude that the HSC does not have any strong preference in $w$CDM model over $\Lambda$CDM.

Fig.~\ref{fig:data-2d-hsc-x-des} shows the joint posterior of this paper and a similar analysis using 2PCF and aperture-mass skewness applied on DES-Y3 data (Gomes et al. in prep.). Here, we assumed the independence of two data sets because the footprint overlap is not large. The joint posterior is obtained just by multiplying the two marginal posteriors in $(\Omega_{\rm m}, S_8)$ space from HSC-Y3 and DES-Y3. The two posteriors have similar constraining power and therefore the combined posterior increases the FoM by a factor of about two. The tension of the joint posterior to Planck-2018 CMB is found to be $3\sigma$.

\subsection{Validation with mock data}\label{sec:mock-result}
\begin{figure*}[t]
    \centering
    \includegraphics[width=\linewidth]{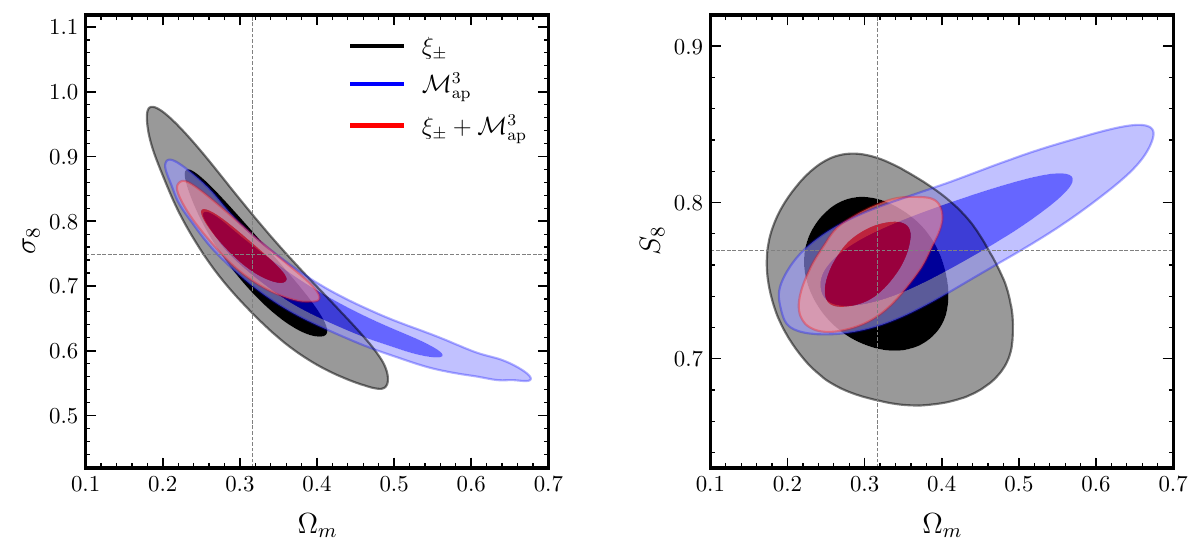}
    \includegraphics[width=\linewidth]{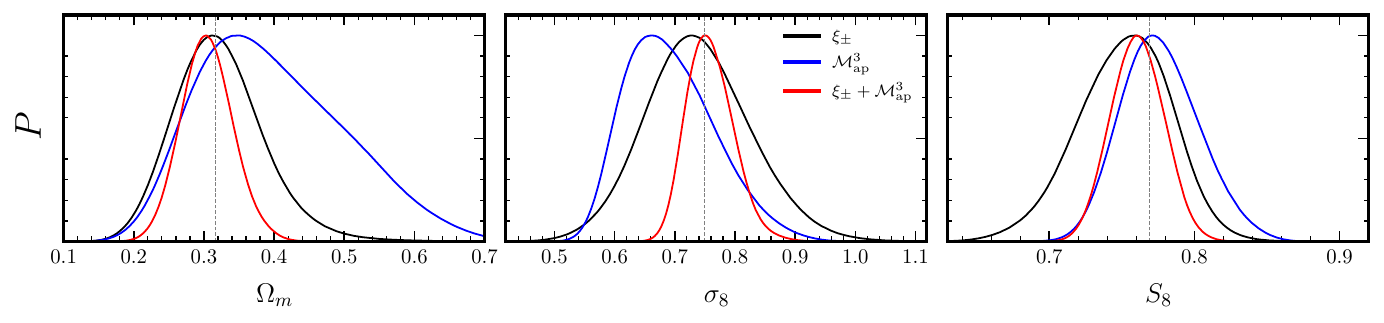}
    \caption{Same as Fig.~\ref{fig:data-2d-1d-omsig8-oms8}, but the analysis results on the fiducial mock data vector. The input model parameters for the simulated mock data vector are indicated by the horizontal/vertical dashed lines. All the recovered posteriors lie withing $1\sigma$ of the input cosmological parameters  which validates the robustness of 
    the analysis pipeline and analysis choices adopted in this paper. A slight bias in $(\Omega_{\rm m}, S_8)$-posterior from the aperture-mass-skewness-only analysis (the blue contour) shown in the upper right panel is due to the prior volume effect and parameter degeneracy between $\Omega_{\rm m}$ and $\Delta z_{2,3}$, which can be seen in Fig.~\ref{fig:mock-triangle}. If we adopt a 3\%   prior on $\Delta z_{2,3}$, this bias is reduced as demonstrated in Fig.~\ref{fig:mock-2d-omsig8-oms8-dz23-prior}.}
    \label{fig:mock-2d-1d-omsig8-oms8}
\end{figure*}

\begin{figure*}[t]
    \centering
    \includegraphics[width=\linewidth]{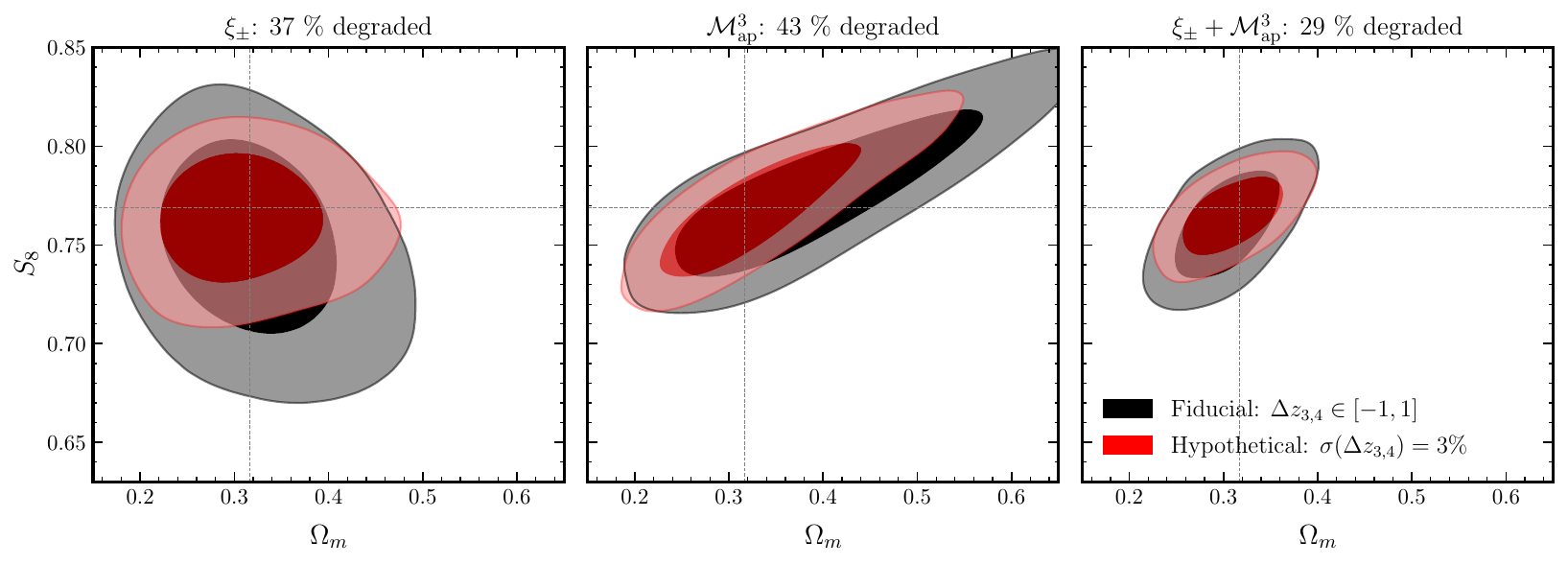}
    \caption{Comparison of the mock analysis results from the fiducial setup (black) and hypothetical setup (red). From left to right panels, the posteriors from the 2PCF-only, the aperture-mass-skewness-only, and the joint analyses are shown. In the fiducial setup, we adopt an uninformative prior on the residual photo-z bias parameter $\Delta z_{1,2}$, while in the hypothetical setup we adopt a 3\% Gaussian prior on them. Note that the fiducial result shown in black is identical to that shown in the same color in Fig.~\ref{fig:mock-2d-1d-omsig8-oms8}. The level of degradations in terms of FoM compared to the hypothetical setup are indicated in the title of each panel. Note also that for the middle panel, from $\mapcube{}{}$ alone, there is a small bias in in the fiducial analysis. }
    \label{fig:mock-2d-omsig8-oms8-dz23-prior}
\end{figure*}

\begin{figure*}[t]
    \centering
    \includegraphics[width=\linewidth]{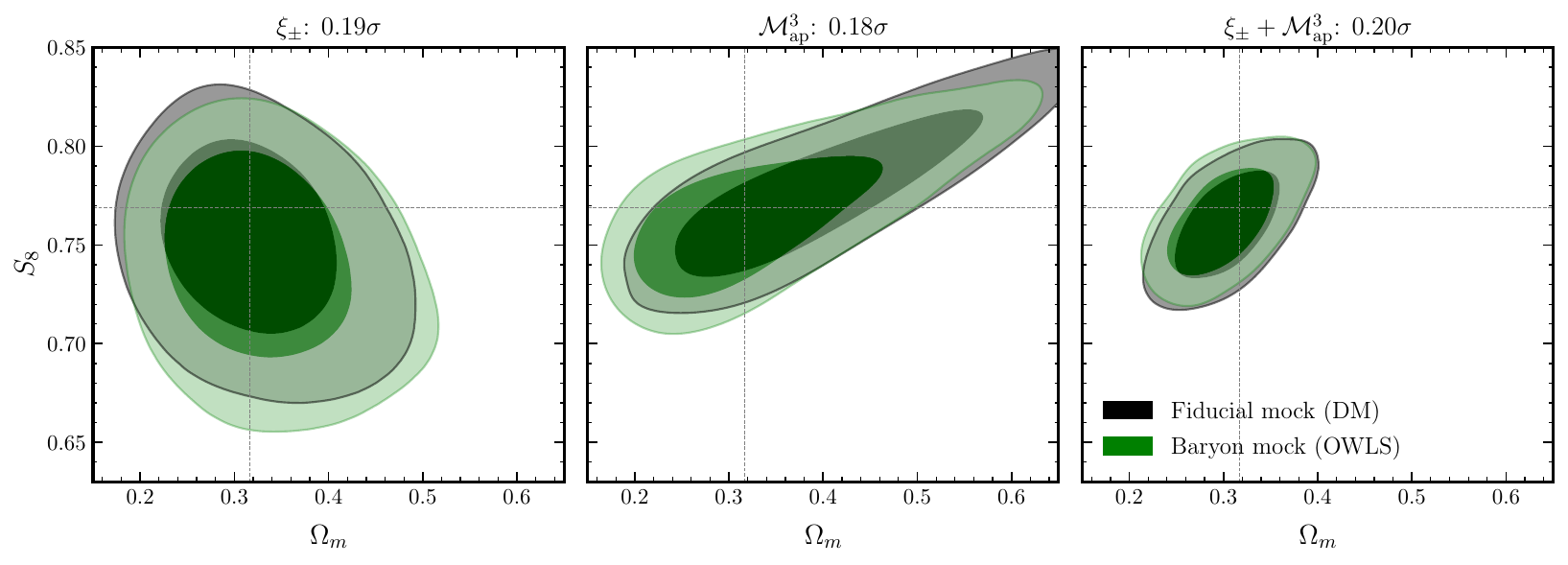}
    \caption{Validation  of the analysis setup against  baryonic effects on the data vector. The black contour is the analysis result on the fiducial mock data vector that is simulated without any baryonic effect on matter power spectrum and bispectrum, while the green contour is the result on the baryon-contaminated mock data vector with baryonic feedback as large as OWLS simulation. The level of bias in the cosmological parameters $\Omega_{\rm m}$ and $S_8$ is indicated in the title for each panel with and without the baryonic effect.}
    \label{fig:mock-2d-oms8-baryon-owls}
\end{figure*}

In this section, we present the result of the validation of the analysis pipeline and analysis choices, through the mock analysis. We generate the fiducial mock data vector by using our analysis pipeline with the following input model parameters; $S_8=0.769$, $\Omega_{\rm m}=0.3166$, $n_{\rm s}=0.9649$, $h_0=0.6727$, $\omega_{\rm b}=0.02236$, $w=-1$, $A_{\rm IA}=0.0$, $\alpha_{\rm IA}=1.0$, $\Delta z_{1,2}=0.0$, $\Delta z_{3}=0.11$, $\Delta z_4=0.19$, $m_{1,2,3,4}=0.0$ and $\alpha'^{(2)}=\alpha'^{(4)}=\beta'^{(2)}=\beta'^{(4)}=0$.

The result of the analysis on the fiducial mock data vector is shown in Fig.~\ref{fig:mock-2d-1d-omsig8-oms8}. 
In order to assess the bias of posterior quantitatively, we define the metric of the bias in the following way; we first compute the total probability that has a smaller probability density than the input cosmological parameter (which intuitively corresponds to $p$-value)
\begin{align}
    p = \int_{\mathcal{P}(\Omega_{\rm m}, S_8)<\mathcal{P}(\Omega_{\rm m}^{\rm in}, S_8^{\rm in})}{\rm d}\Omega_{\rm m}{\rm d}S_8~
    \mathcal{P}(\Omega_{\rm m}, S_8) \ .
\end{align}
Throughout this section, we focus on the bias in $(\Omega_{\rm m}, S_8)$, as these are the primary parameters of interest.
The probability in the above equation can be easily estimated using the API of \code{getdist} since this is only 2D posterior distribution without relying on any machine-learning-based density estimator that is typically needed for the analysis of high dimensional posterior distribution. Then we translate the probability in terms of the 1D normal distribution, $n\equiv \sqrt{2}{\rm Erf}^{-1}(p)$, leading to ``$n$--$\sigma$'' expression for the bias quantification.

We find that the biases in the 2PCF-only, the aperture-mass-skewness-only, and the joint analyses are 
$0.07\sigma$, $0.28\sigma$, and $0.16\sigma$, respectively. 
The relatively large bias in the aperture-mass-skewness-only analysis is due to the parameter degeneracy between $\Omega_{\rm m}$ and $\Delta z_{3,4}$, and the prior volume effect. This bias can be alleviated by introducing a hypothetical tight prior on $\Delta z_{3,4}$ as demonstrated in Fig.~\ref{fig:mock-2d-omsig8-oms8-dz23-prior}, although such a tight prior is not practically available for now because it requires much more reliable photo-$z$ calibration at high redshift than the current one (for this reason we call this experiment ``hypothetical''). 
Although this is a hypothetical situation at the time of this paper, with the photo-$z$ calibration with more high redshift spectroscopic survey data, e.g. DESI, this may be accomplished in real data.
We also note that the degradation of the FoM due to the uninformative $\Delta z_{3,4}$ prior is only 29\% in the joint analysis, thanks to the self-calibration of the photo-$z$ bias through the scale dependence and cross-correlation with lower redshift bins which is accurately and reliably calibrated. We leave further detail of the degeneracy between the cosmological parameters and the photo-$z$ bias parameters in Fig.~\ref{fig:mock-triangle} in Appendix~\ref{sec:large-triangle-contour}. Despite this relatively large bias in the aperture-mass-skewness-only analysis, the result from the joint analysis has much smaller bias, because the prior volume effect that causes the apparent bias is alleviated by combining the 2PCF and the aperture-mass skewness.

To validate the scale cut of aperture-mass skewness, we generate a baryon-contaminated mock data vector that incorporates baryonic feedback effect on the spatial matter density distribution as large as OWLS simulation. For a consistent treatment of the baryonic feedback effect in 2PCF and aperture-mass skewness, we follow the prescription described in \citet{Gomes.Weller.2025}. For 2PCF, we multiply the scale-dependent suppression factor measured from OWLS simulation on the matter power spectrum using the \code{baryon\_power\_scaling} module in \code{cosmosis-standard-library}. For aperture-mass skewness, we multiply the scale-dependent suppression factor measured from TNG-300 in \citet{Takahashi.Shirasaki.2019} to the matter bispectrum in Eq.~(\ref{eq:3pcf-theory}), while the overall amplitude of the suppression factor is rescaled by a factor of $1.5^{3/2}=1.84$ so that the peak suppression factor matches the peak suppression factor that would be applied in the OWLS simulation. The rescaling factor is motivated by the physical insight that the suppression due to baryonic effect is effective at the 1-halo regime, in which we expect the scaling relation of $B_{\rm m}(k)\propto u^3(k) \propto [P(k)]^{3/2}$ where $u(k)$ is the halo density profile, and by the observation that the peak suppression factor to power spectrum in OWLS is larger than the TNG-300 scenario by a factor of $1.5$.

The result of the analysis on the baryon-contaminated mock data vector is presented in Fig.~\ref{fig:mock-2d-oms8-baryon-owls}. 
Note that here the model assumes the dark-matter-only power and bi spectra, and the identical analysis choice is applied to both of the fiducial and the baryon-contaminated data vectors. 
In the 2PCF-only analysis, the bias increases to $0.19\sigma$ when the data vector has a baryonic effect on the power spectrum as large as in the OWLS simulation. In the aperture-mass-skewness-only analysis, the bias reduces to $0.18\sigma$. This is because of the net effect of the prior-volume effect as seen in Fig.~\ref{fig:mock-2d-omsig8-oms8-dz23-prior} and the suppression due to baryon, and therefore this is just coincidence for this survey and does not indicate any physical cancellation of two different types of biases. 
However, once they are combined in the joint analysis, we find that the bias is significantly suppressed up to $0.20\sigma$, and the result is stable regardless of the presence of the baryonic effect in the data vector.
In \citet{Gomes.Weller.2025}, we found that aperture-mass-skewness-only analysis biases $S_8$ higher in the presence of the baryonic effect because the scale cut we used for the DES data accidentally corresponds to the scale where the baryonic effect causes enhancement in bispectrum, and that the bias cancels when combined with 2PCF-only analysis that biases $S_8$ lower. 
In this paper, we include the aperture-mass skewness signal down to a relatively smaller scale than \citet{Gomes.Weller.2025}, and therefore we found that $S_8$ is biased lower in aperture-mass-skewness-only analysis as well because our minimum scale is probing a smaller scale where suppression becomes dominant than the enhancement seen in the paper. 
Despite these biases in the 2PCF-only and the aperture-mass-skewness-only analyses in the same direction (lower $S_8$), it is interesting that the joint analysis provides a smaller bias in $S_8$. 
This is thanks to degeneracy breaking and the reduced prior-volume effect in the joint analysis. More concretely, as can be seen in the left and middle panels of Fig.~\ref{fig:mock-2d-oms8-baryon-owls}, the biases for the 2PCF-only and the aperture-mass-skewness-only analysis is driven by the posterior volume at different locations (around $(\Omega_{\rm m}, S_8)\sim(0.3, 0.68)$ for 2PCF and $(0.24, 0.0.73)$ for aperture-mass skewness, respectively), and therefore the biases are suppressed when they are combined.
However, we also note that here the bias is assessed in the unit of statistical error of the fiducial analysis setup. If we employ the tight prior on the photo-$z$ parameters as in Fig.~\ref{fig:mock-2d-omsig8-oms8-dz23-prior}, the constraining power increases and the relative bias increases to $0.34\sigma$, and therefore the appropriate baryonic model is needed to marginalize over the baryonic systematic uncertainty.

\section{Conclusion}\label{sec:conclusion}
In this work, we conducted a joint cosmological analysis using the two-point cosmic shear correlation functions and the aperture-mass skewness measured from the Year 3 data of the Hyper Suprime-Cam Subaru Strategic Program (HSC-Y3). This study represents the first cosmological application of three point correlations to HSC data and explores the added value of incorporating non-Gaussian information in weak lensing analyses.

The aperture-mass skewness, a compressed statistic of the three-point correlation function, was adopted to efficiently extract essentially all the third-order information. Given the large dimensionality of the original data vector, we applied data compression techniques to feasibly extract cosmological information. The joint analysis of the 2PCF and aperture-mass skewness improved constraints in the $S_8$–$\Omega_m$ plane, yielding an 80\% gain in the figure of merit (FoM) relative to the 2PCF-only analysis. This improvement is attributed to the breaking of parameter degeneracies by the higher-order statistic. 
We found no evidence for intrinsic alignment in the aperture-mass skewness signal, and the two- and three-point statistics were found to be internally consistent across redshift bins and scales used in this paper. Notably, the inclusion of aperture-mass skewness led to a slightly lower value of $S_8$, which increased the tension with the Planck 2018 results to 3.2$\sigma$ in the $S_8$–$\Omega_m$ space.
We also explored constraints in the extended $w$CDM model but found no significant deviation from $w=-1$, indicating no strong preference for dynamical dark energy with the current data.

We explored the improvement in handling systematic errors with the combination of two- and three-point functions. We find, as noted by \citet{Gomes.Weller.2025}, that the joint constraints have lower bias in the presence of baryonic feedback effects. We also find improved performance given the uncertainties in the redshift distributions of source galaxies. Our constraints are based on uninformative priors on the third and fourth redshift bin, with the joint analysis leading to improved and robust posteriors on both cosmological and nuisance parameters.  

These results demonstrate the scientific value of higher-order statistics and data compression techniques in weak lensing cosmology. Our work highlights the feasibility and benefits of including compressed non-Gaussian information and paves the way for similar approaches to be applied in forthcoming Stage-IV surveys such as LSST, Euclid and Roman.

\section*{Acknowledgment}
We thanks Mike Jarvis, Minsu Park, Kazuyuki Akitsu, Tianqing Zhang, Rachel Mandelbaum, and Xiangchong Li for useful discussion. We also thank Masahiro Takada for reviewing this paper. SS is supported by the JSPS Overseas Research Fellowships, and BJ and RCHG are partially supported by the US Department of Energy grant DE-SC0007901.

\bibliography{refs}

\appendix
\section{Null test}\label{sec:null-test}
\begin{figure*}[t]
    \centering
    \includegraphics[width=\linewidth]{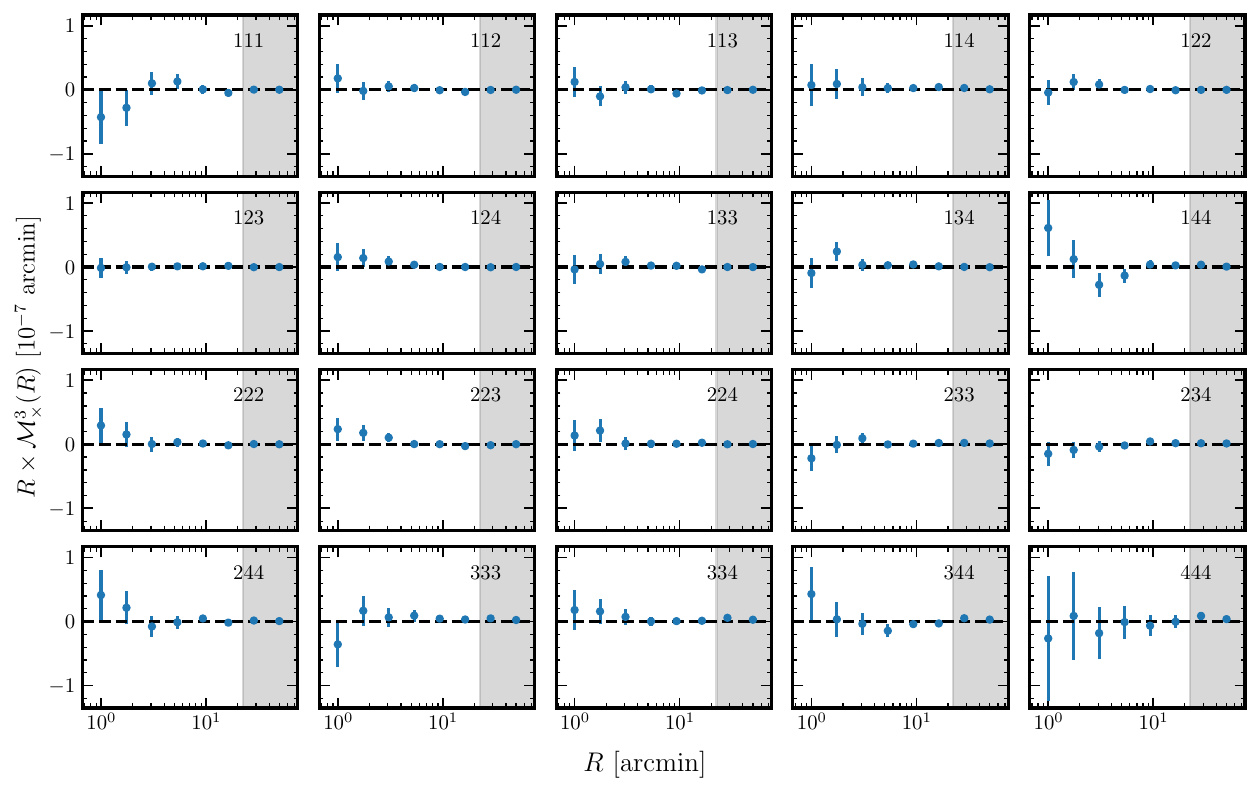}
    \caption{The result of the null test of the mass skewness signal. Here, test is done on one of the B-mode signals of mass skewness $\mathcal{M}_\times^3$. The significance of the B-mode is $\chi^2=9.0$ for a degree of freedom ${\rm dof}=120$, yielding the $p$-value for the null hypothesis $p=0.99$.}
    \label{fig:signal-mx3}
\end{figure*}

In this Appendix, we present the result of the null test of the third-order shear statistics, namely $\mathcal{M}_\times^3$. It is known that the weak lensing signal yield no cross signal $\mathcal{M}_\times^3$, and therefore it is used to test the observational systematic effect in the associated mass aperture skewness signal $\mathcal{M}_{\rm ap}^3$. The measurement result of the cross signal is shown in Fig.~\ref{fig:signal-mx3}. The chi-squared value for this cross signal from zero is found to be $\chi^2=9.0$ for degree-of-freedom ${\rm dof}=120$, yielding the $p$-value for the null hypothesis $p=0.99$. Therefore we conclude that our mass aperture skewness signal is not contaminated by the significant systematics effect.

\section{mass aperture skewness emulator}\label{sec:map3-emulator}
\begin{table}
\caption{Parameter ranges of model parameters used for emulator training.}
\label{tab:emu-range}
\setlength{\tabcolsep}{15pt}
\begin{center}
\begin{tabular}{ll}
\hline
Parameter & Range \\ \hline
$S_8$                   & $(0.6, 0.95)$\\
$\Omega_{\rm m}$        & $(0.1,0.5)$\\
$n_{\rm s}$             & $(0.87,1.07)$\\
$h_0$                   & $(0.62,0.80)$\\
$\omega_{\rm b}$        & $(0.02, 0.025)$\\
$w$                     & $(-2.0, -0.5)$\\
\hline
\end{tabular}
\end{center}
\end{table}

\begin{figure*}[t]
    \centering
    \includegraphics[width=\linewidth]{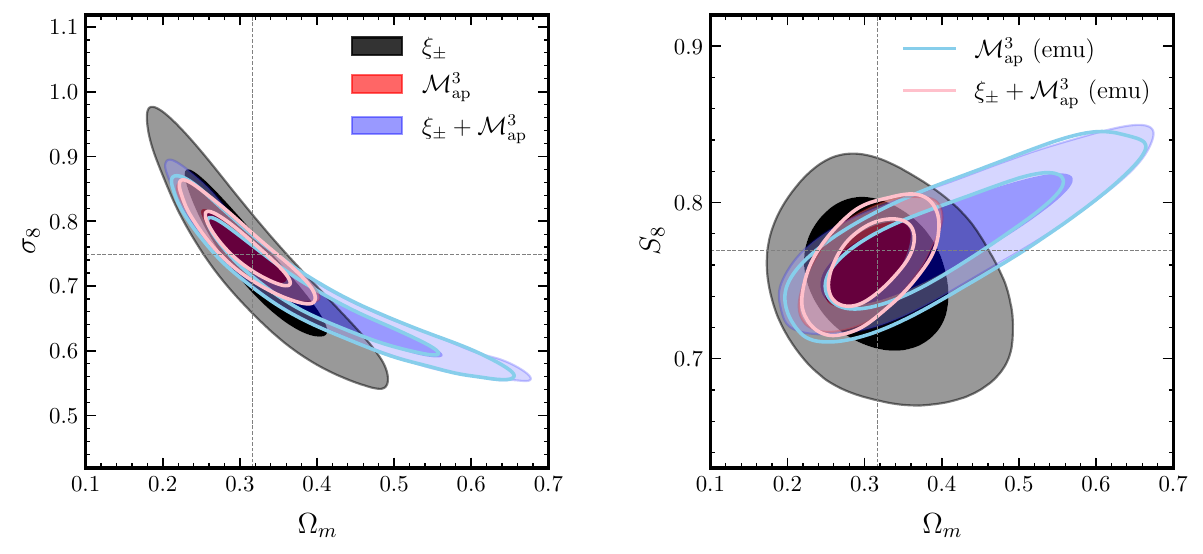}
    \caption{The validation of the mass-skewness emulator. The filled contours are the analysis results where we use the mass-skewness emulator for inference on the mass-skewness mock data vector generated using fastnc, while the unfilled contours are the analysis results on the mock data vector generated by the same mass-skewness emulator as used for inference. The perfect overlap of the filled and unfilled contours indicates the robustness of the mass-skewness emulator.}
    \label{fig:mock-2d-omsig8-oms8-emuvalidate}
\end{figure*}

For efficient parameter inference, we train an emulator of mass aperture skewness. In order to reduce the dimension of the input model parameters for the emulator, we train the mass-skewness without line-of-sight integration by following \citet{Gomes.Weller.2025}. We define the redshift-dependent mass aperture skewness by
\begin{align}
    \mathcal{M}_{ap}^{3,abc}(R) \equiv \int_0^{\chi_{\rm H}} {\rm d}\chi 
    \frac{q_a(\chi)q_b(\chi)q_c(\chi)}{\chi^4} \mathcal{M}_{\rm ap}^3\left(R, z(\chi)\right) \ .
\end{align}
By separating out the lensing kernel from the mass aperture skewness, the resultant redshift-dependent mass aperture skewness $\mathcal{M}_{\rm ap}^3(R,z)$ is independent of the survey-dependent parameters such as $\Delta z_a$, and therefore we can simplify the design of the emulator.

In order to train the emulator, we draw the 20,000 Latin Hyper Cube (LHC) samples of model parameters from ranges summarized in Table~\ref{tab:emu-range}. For each LHC sample, we compute the redshift-dependent mass aperture skewness by the multipole-based method using \code{fastnc}. For the computation of redshift-dependent mass aperture skewness, we use the same 8 filter radii as in the measurement, and the 130 redshift points (logarithmically spaced in $[10^{-4}, 10^{-1}]$ and linearly spaced in $[0.11, 3]$), yielding in total 1040 data points for each LHC sample. We discard samples that failed the computation due to the outlier model parameters. We divide the samples into 80\%, 10\%, and 10\% sub samples for training, validating, and testing the emulator, respectively. We build the emulator using the \code{tensorflow keras}. Our emulator consists of two networks; the former emulates the first 80 principle components (PCs) and the latter emulates the residual compared to the prediction by the PCs, which we call PCANet and ResNet, respectively. The PCANet emulates the nonlinear relation between the input model parameters and the output PCs of data, and therefore we use a deep and narrow neural network with hidden dimensions of $[126, 256, 256, 126]$, while the ResNet emulates the details of the residuals that has high degree-of freedom and therefore we use a relatively wide neural network with hidden dimensions of $[256,256,256,256]$. After training, we convert the \code{keras} model to the \code{tflite} model to reduce the model size while keeping the performance.

We test the performance of the emulator at the level of parameter inference. In Fig.~\ref{fig:mock-2d-omsig8-oms8-emuvalidate}, we show the results on the mock data generated by \code{fastnc}, and the mock data generated by the emulator. Note that we used the emulator for parameter inference in both cases. From the perfect overlap of the posteriors on mocks by \code{fastnc} and emulator, we conclude the robustness of emulator.

\section{Low $\Omega_{\rm m}$ makes a skewed posterior}\label{sec:low-om-mock}
\begin{figure}[t]
    \centering
    \includegraphics[width=\linewidth]{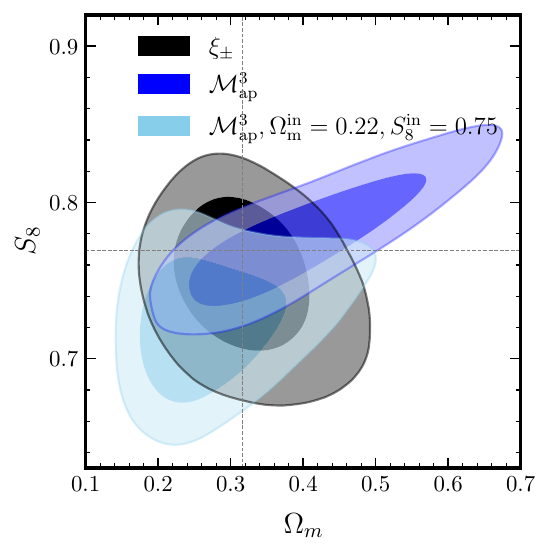}
    \caption{The result of the mass-skewness-only analysis on the mock data where the input cosmological parameter is set to smaller values than the fiducial mock, $(\Omega_{\rm m}, S_8)=(0.22, 0.75)$, shown by the skyblue contour compared to the fiducial mock results. We can see that the posterior can be skewed because of the difference in the underlying cosmological parameters. This explains why the mass-skewness-only posterior from data is skewed in Fig.~\ref{fig:data-2d-1d-omsig8-oms8}.}
    \label{fig:mock-2d-oms8-om02}
\end{figure}
In this Appendix, we show that the low $\Omega_{\rm m}$ preference makes the posterior distribution from mass-skewness-only analysis skewed. We generated the mock data vector that has a low $\Omega_{\rm m} (=0.22)$ as input while other cosmological parameters are fixed to the same as in the fiducial mock data vector that is used in Section~\ref{sec:mock-result}. The estimated posterior distribution from this mock data vector is shown in Fig.~\ref{fig:mock-2d-oms8-om02}, in which the posterior is compared to those estimated from the fiducial mock data vector. We observe that the resultant posterior distribution is skewed and the way to overlap with the posterior from 2PCF changes. This explains why the posterior distribution from the HSC-Y3 mass-skewness-only analysis is skewed and why it has a different degeneracy direction that was found in the fiducial mock analysis (e.g. Fig.~\ref{fig:mock-2d-1d-omsig8-oms8}). The mass aperture skewness measured from HSC-Y3 data prefers the low $\Omega_{\rm m}$, plausibly due to the statistical scatter, and therefore the posterior from it is skewed as demonstrated by the low-$\Omega_{\rm m}$ mock in Fig.~\ref{fig:mock-2d-oms8-om02}.

\section{Large triangle contour}\label{sec:large-triangle-contour}
In this section, we show a comprehensive contour plot that includes more model parameters.
\begin{figure*}[t]
    \centering
    \includegraphics[width=\linewidth]{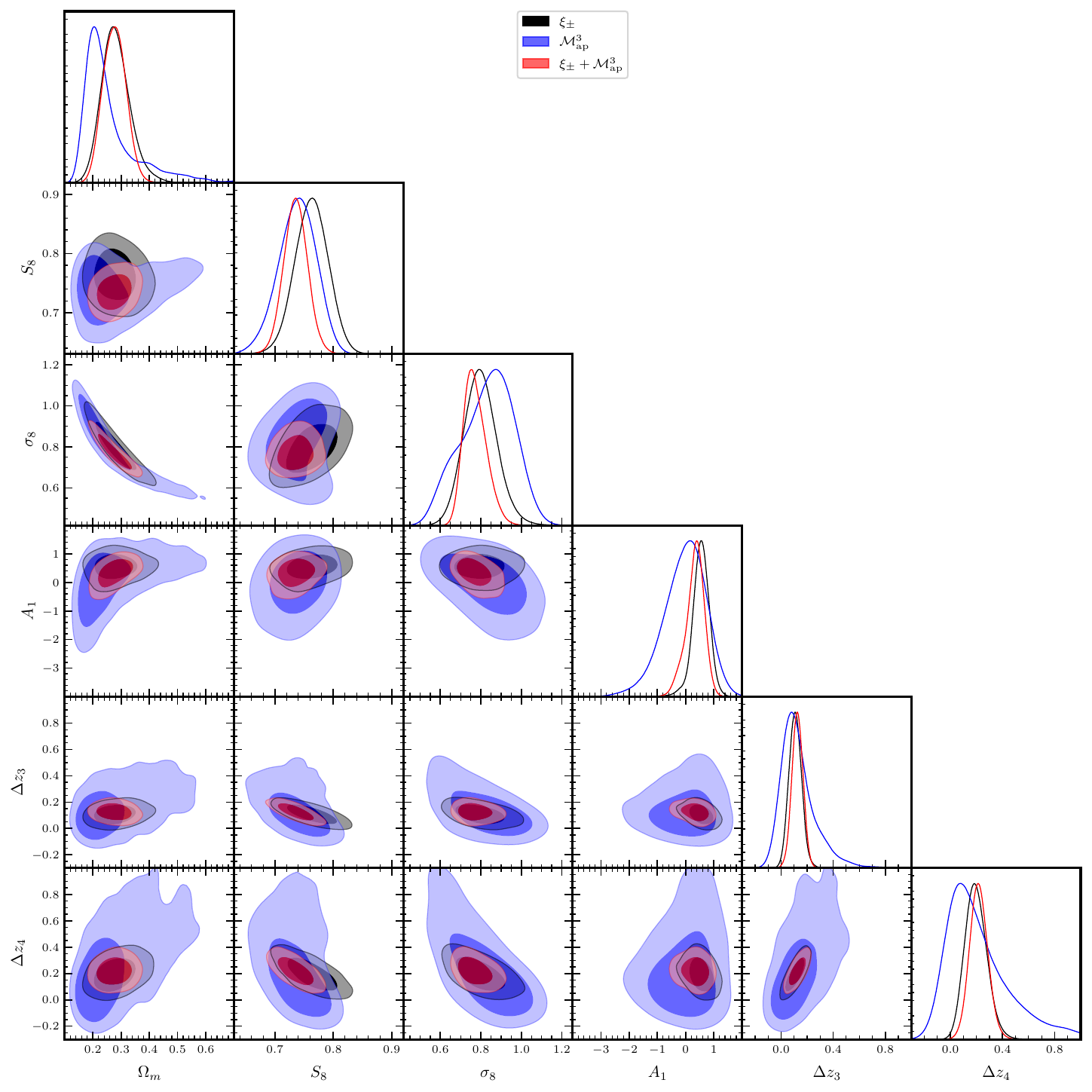}
    \caption{The parameter constraint from the fiducial analyses in this paper, but with more model parameters shown together to present how they degenerate together.}
    \label{fig:data-triangle}
\end{figure*}

\begin{figure*}[t]
    \centering
    \includegraphics[width=\linewidth]{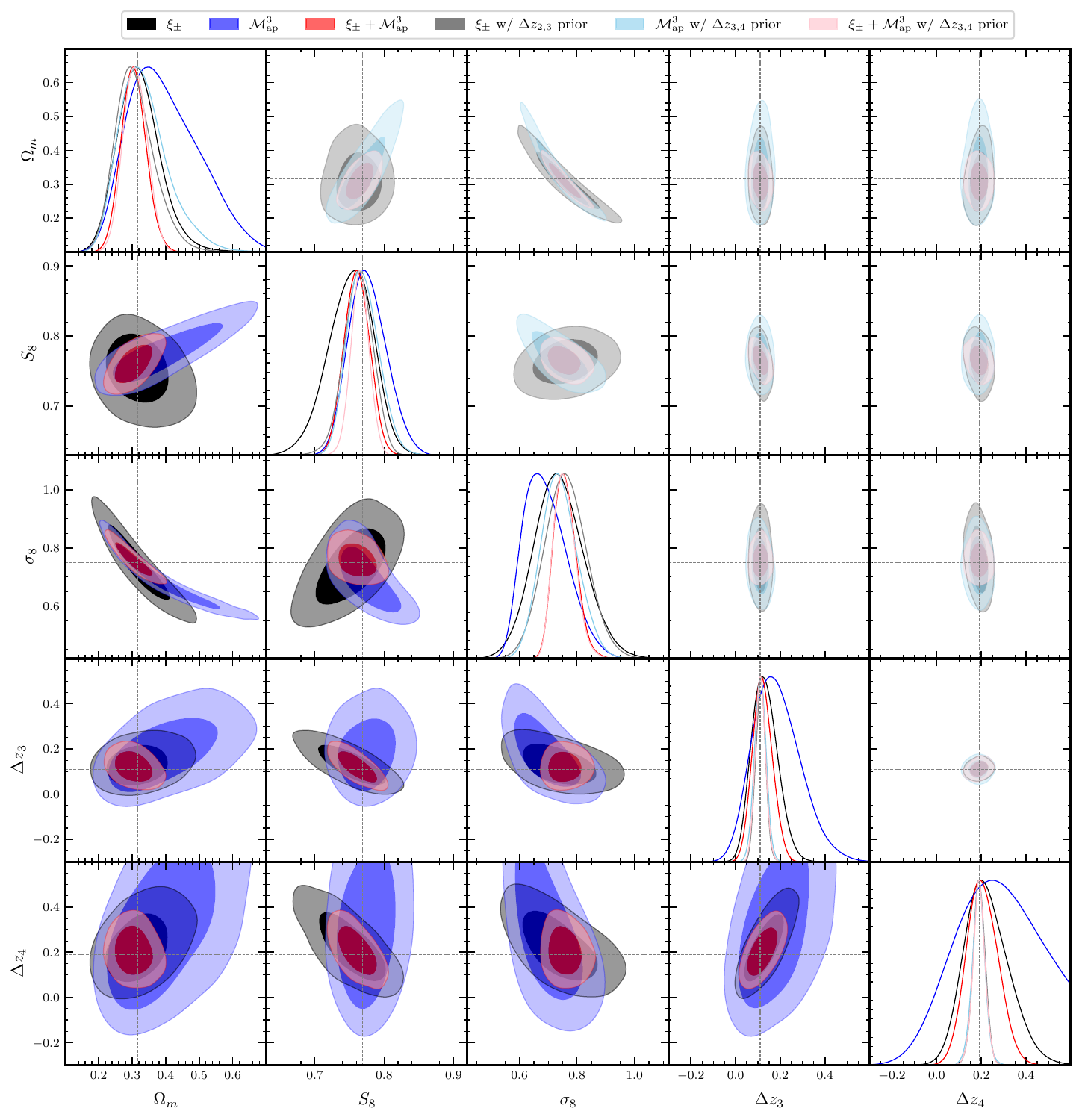}
    \caption{The analysis result on the fiducial mock data. The lower left triangle shows the result from the fiducial analysis where we adopt the uninformative prior on $\Delta z_{3,4}$, while the upper right triangle shows the result from the hypothetical analysis setup where we adopt a 3\% level tight prior on $\Delta z_{3,4}$. In the fiducial analysis setup, we see that $\Delta z_{3,4}$ parameter degenerate with the cosmological parameter, which consequently causes a bias due to the prior-volume effect on the marginalized posterior distribution.}
    \label{fig:mock-triangle}
\end{figure*}

\end{document}